\documentclass[aps,prl,superscriptaddress,twocolumn,longbibliography,floatfix,10pt]{revtex4-1}

\usepackage{orcidlink}
\usepackage{bm}
\usepackage{bbm}
\usepackage{float}
\usepackage{graphicx}
\usepackage{amsmath}    
\usepackage{comment}
\usepackage{units}
\usepackage{amsthm}
\usepackage{amssymb}
\usepackage{bbold}
\usepackage{xcolor}
\usepackage{amssymb}
\usepackage{booktabs}
\usepackage{ulem}
\usepackage{multirow}

\usepackage{hyperref}
\hypersetup{breaklinks, colorlinks=true, urlcolor=blue,linkcolor=blue, anchorcolor=blue, citecolor=blue}
\newcommand{\ket}[1]{|#1 \rangle}
\newcommand{\braket}[2]{\langle #1|#2\rangle}
\newcommand{\ketbra}[2]{|#1\rangle\langle#2|}
\newcommand{\mel}[3]{\langle #1 \vert #2 \vert #3\rangle}

\newcommand{\tr}[1]{\text{Tr}\left(#1\right)}

\renewcommand{\eqref}[1]{Eq.~(\ref{#1})} 
\newcommand{\figref}[1]{Fig.~\ref{#1}} 

\begin{document}

\title{Exponentially faster preparation of quantum dimers via driven-dissipative stabilization}
\author{Kian Hwee Lim \orcidlink{0000-0003-2154-4288}}
\thanks{Equal contribution}
\email{kianhwee\_lim@u.nus.edu}
\affiliation{Centre for Quantum Technologies, National University of Singapore, 3 Science Drive 2, Singapore 117543}
\author{Wai-Keong Mok \orcidlink{0000-0002-1920-5407}}	
\thanks{Equal contribution}
\affiliation{Institute for Quantum Information and Matter, California Institute of Technology, Pasadena, CA 91125, USA}

\author{Jia-Bin You\orcidlink{0000-0001-8815-1855}}
\affiliation{Institute of High Performance Computing, A*STAR (Agency for Science, Technology and Research), 1 Fusionopolis Way, \#16-16 Connexis, Singapore 138632}
\author{Jian Feng Kong\orcidlink{0000-0001-5980-4140}}
\affiliation{Institute of High Performance Computing, A*STAR (Agency for Science, Technology and Research), 1 Fusionopolis Way, \#16-16 Connexis, Singapore 138632}

\author{Davit Aghamalyan \orcidlink{0009-0007-4926-9739}}
\affiliation{Institute of High Performance Computing, A*STAR (Agency for Science, Technology and Research), 1 Fusionopolis Way, \#16-16 Connexis, Singapore 138632}

\begin{abstract}
We propose a novel rapid, high-fidelity, and noise-resistant scheme
to generate many-body entanglement between multiple qubits
stabilized by dissipation into a 1D bath. Using a carefully
designed time-dependent drive, our scheme achieves a provably exponential
speedup over state-of-the-art dissipative stabilization schemes in $1$D
baths, which require a timescale that diverges as the target
fidelity approaches unity and scales exponentially with the number of
qubits. To prepare quantum dimer pairs, our scheme only requires
local $2$-qubit control Hamiltonians, with a protocol time that is
independent of system size. This provides a scalable and robust
protocol for generating a large number of entangled dimer pairs on-demand, serving as a
fundamental resource for many quantum metrology and quantum information
processing tasks.
\end{abstract}

\maketitle

\paragraph{Introduction.---}
Entangled quantum states are essential for quantum
computation~\cite{nielsen2002quantum} and
metrology~\cite{giovannetti2011advances}, which demand their high fidelity
generation in a way that is resilient to noise and dissipation. Dissipation,
once seen as detrimental, is now explored as a resource for entanglement
generation~\cite{kastoryano2011dissipative}. However, despite a plethora of
theoretical proposals and experimental realizations for generating entangled
states with cavity quantum electrodynamics (QED)
systems~\cite{kastoryano2011dissipative,reiter2012driving,sweke2013dissipative,su2014scheme,shen2011steady},
ion traps~\cite{lin2013dissipative,cole2021dissipative,cole2022resource},
Rydberg
atoms~\cite{li2020periodically,shao2014stationary,chen2018accelerated,rao2014deterministic},
colour
centers~\cite{qiao2020phononic,jin2019preparation,rao2017dissipative,li2012dissipative},
circuit QED~\cite{leghtas2013stabilizing,reiter2013steady}, optical lattices
and spin
chains~\cite{ramos2014quantum,kordas2012dissipation,botzung2021engineered,morigi2015dissipative,de2017steady},
limitations persist in either the speed of state generation, entanglement
fidelity or the aforementioned robustness to noise and dissipation. For
instance, the dissipative entanglement generation schemes based on
Ref.~\cite{kastoryano2011dissipative} rely on perturbative expansions in the
system's driving strengths, which fundamentally limits the speed of
entanglement generation.

It was also shown in~\cite{pichlerQuantum2015,ramosNon2016} that when
multiple locally-driven system qubits are coupled to a chiral 1D bath (which
could either be a waveguide or a spin chain), one can obtain many-body
entangled states stabilized by the dissipation into the 1D bath. In this
theoretical scheme, no perturbative expansions in the system's driving
strengths are required, which circumvents the aforementioned speed limit. An
atomic implementation of this scheme on cold quantum gases was proposed
in~\cite{ramos2014quantum}, and experimentally implemented recently on
superconducting qubits~\cite{shah2024stabilizing}.

However, as we will demonstrate in this manuscript,
time-independent many-body entanglement generation schemes using engineered
dissipation as proposed
in~\cite{pichlerQuantum2015,ramosNon2016,ramos2014quantum} require a
timescale that diverges as the target fidelity approaches unity, leading to
an inevitable tradeoff between fidelity and speed. Furthermore, for existing
steady state schemes including~\cite{gutierrez2023dissipative}, the
protocol time scales exponentially with the number of qubits. This presents a
severe limitation for scaling up to many qubits, especially in the presence
of noise. We propose a new scalable protocol based on
carefully designed time-dependent driving to generate many-body entanglement
in 1D systems in a fast, high-fidelity and noise-robust manner. To the
best of our knowledge, our scheme is the only one that fulfils this
trifecta.

An important application of our scheme is in preparing a large number of
quantum dimer pairs on-demand, which are valuable resource states for various quantum
technologies such as quantum
metrology~\cite{gutierrez2023dissipative,groszkowski2022reservoir} and
quantum information processing. Our scheme achieves a high-fidelity
preparation using only local $2$-qubit control Hamiltonians, rendering it
feasible to current experimental capabilities. Crucially,
our protocol time is independent of the number of qubits, thereby
exponentially faster than the previously proposed
schemes~\cite{ramos2014quantum,pichlerQuantum2015,ramosNon2016,gutierrez2023dissipative}.
We perform a systematic study of robustness of our scheme against various
sources of noise and decoherence. We show that in the presence of
any amount of spontaneous decay outside of the 1D bath, previous
time-independent schemes eventually fail for a sufficiently large number of
qubits due to the exponentially long timescales required. On the contrary,
our scheme is robust against such losses for any number of qubits.

\paragraph{Many-body entangled dark states of 1D systems. ---}
\begin{figure}
  \centering
  \includegraphics[width=0.47\textwidth, trim={0cm 0cm 0cm 0cm},clip]{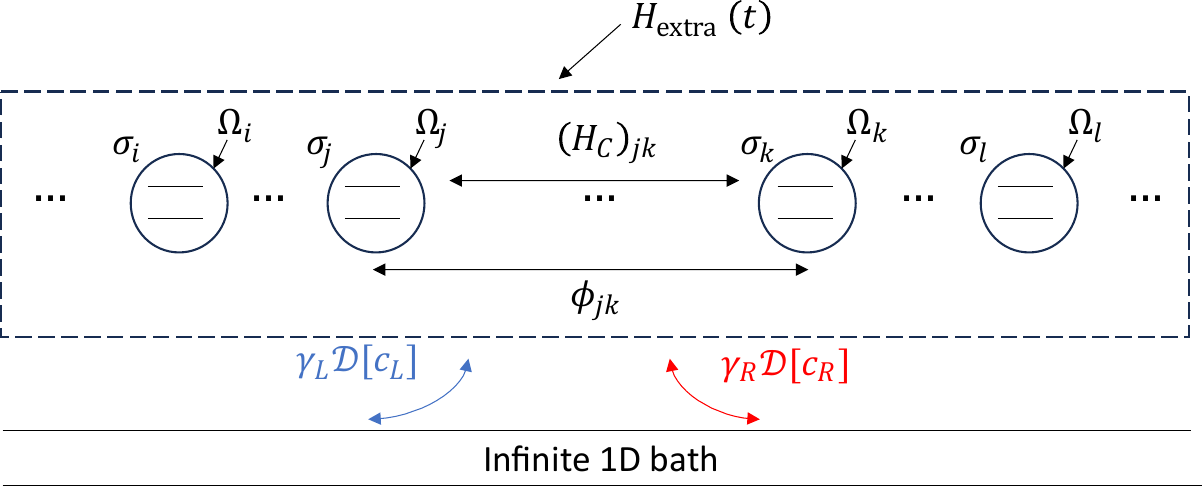}
  \caption[]{Schematic for the setup described by \eqref{eqn: full
  hamiltonian} and \eqref{eqn: main master eqn}. $N$ qubits are coupled to a
  waveguide as per~\cite{pichlerQuantum2015} or to a 1D spin chain with a
  synthetic gauge field as per~\cite{ramosNon2016}. Here, $\sigma_j$ is the
  lowering operator for $j$th system qubit that is driven with an external
  local drive $\Omega_j$. $\phi_{j k}$ describes the phase picked up by the
  bath excitation as it travels between the $j$th and $k$th qubit along the
  infinite 1D bath, which affects the bath-mediated chiral interaction
  between the $j$th and $k$th qubit. $H_\text{extra}$ is the extra external
  field in our scheme which we will introduce later. All the qubits decay
  collectively into the 1D bath through collective jump operators $c_L$
  (left-going modes) and $c_R$ (right-going modes) with decay rates
  $\gamma_L$ and $\gamma_R$ respectively.}
  \label{fig: prototypical set-up}
\end{figure}
In waveguide QED, one often considers the case where there are $N$ qubits
coupled to a $1$D bath~\cite{castellsAtomic2021,ramosNon2016}. The $1$D bath
serves firstly as a decay channel for the system qubit excitations, and
secondly to mediate long-distance coherent interactions between the system
qubits. With reference to \figref{fig: prototypical set-up}, under the
Born-Markov and rotating wave approximations, by tracing out the $1$D bath,
we obtain the following Hamiltonian (setting $\hbar = 1$) for the $N$ system
qubits
\begin{equation}
  \label{eqn: full hamiltonian}
  H_S = - \sum_{i=1}^N\delta_i \sigma_i^\dag \sigma_i + H_\text{drive}(t) + \sum_{j<k} (H_C)_{j k}
\end{equation}
where 
$(H_C)_{j k} = \frac{i}{2}(\gamma_R e^{-i\phi_{j k}} - \gamma_L e^{i\phi_{j
k}})\sigma_j^\dagger \sigma_k + \text{H.c}$ describes the coherent
interaction mediated by the 1D bath between the $j$th and $k$th system qubits,
$H_\text{drive}(t) = \sum_{i=1}^N (\Omega_i(t)/2) \sigma_i +
\text{H.c}$ describes the local driving on the qubits with Rabi frequency
$\Omega_i(t)$, $\delta_i$ describe the detuning between the $i$th qubit and the carrier frequency of the 1D bath. The dissipation of the system into the 1D bath is described by a
master equation for the $N$ system
qubits~\cite{pichlerQuantum2015,ramosNon2016}
\begin{equation}
  \label{eqn: main master eqn}
  \dot{\rho} = -i [H_S, \rho] + \gamma_L\mathcal{D}[c_{L}]\rho + + \gamma_R\mathcal{D}[c_{R}]\rho.
\end{equation}
Here, $\mathcal{D}[c_{L(R)}]\rho =
c_{L(R)}\rho c_{L(R)}^\dagger - \{c_{L(R)}^\dagger c_{L(R)}, \rho\}/2$
describes the leftward (rightward) dissipation of the system qubits into the
bath, where $c_L = \sum_{j=1}^N e^{i \phi_j}\sigma_j$,
$c_R = \sum_{j=1}^N e^{-i \phi_j}\sigma_j$ are the collective
jump operators. The system is chiral if $\gamma_L \neq \gamma_R$, physically
manifesting as an asymmetric emission into the bath. While entanglement
generation schemes which operate in the transient regime for these $1$D
systems have been proposed~\cite{mok2020microresonators, mok2020long}, a
higher fidelity that is also stabilized by the dissipation into the
bath can be attained in the steady
state~\cite{pichlerQuantum2015,gutierrez2023dissipative}. In particular, it
was shown~\cite{pichlerQuantum2015} that when $\phi_{jk} \text{ mod
}2\pi = 0$, together with certain conditions on $\delta_i$ (or in
the chiral case $\gamma_L \neq \gamma_R$) with homogeneous time-independent
driving $\Omega_i(t) = \Omega$, it is possible to obtain the
following multipartite entangled dark steady state for even $N$
\begin{subequations}
  \label{eqn: most general steady state from zoller}
  \begin{align}
    &\rho_{ss} = \ketbra{\Phi}{\Phi}, \quad \text{where } \ket{\Phi} =
    \prod_{q=1}^{N_m}\ket{M_q} \\
    \ket{M_q} &= a^{(0)}\ket{g}^{\otimes M_q} + \sum_{j_1 < j_2} a^{(1)}_{j_1, j_2}\ket{S}_{j_1 j_2} \ket{g}^{\otimes M_q -2} \nonumber \\
    \label{eqn: multimer definition}
    &+ \dots + \sum a^{(M_q/2)}_{j_1,\dots j_{M_q}} \ket{S}_{j_1 j_2} \dots
    \ket{S}_{j_{M_q-1} j_{M_q}}.
  \end{align}
\end{subequations}
We define $\ket{S}_{ij} =(\ket{e}_i\ket{g}_j - \ket{g}_i\ket{e}_j)/\sqrt{2}$ as a
singlet state (or a dimer pair) between qubits $i$ and $j$.
$\ket{\Phi}$ is a product of $N_m$ adjacent multimers $\ket{M_q}$, and each $\ket{M_q}$ is an entangled state over $M_q$ qubits as defined in \eqref{eqn: multimer definition}, where $M_q$ is an even integer.
Note that the summation in the last line of \eqref{eqn: multimer definition} runs over all different pairings of
qubits $\{(j_1,j_2), \dots (j_{M_q-1}, j_{M_q})\}$ with $j_{k} < j_{k+1}$.
It can also be shown that $a^{(i)} \propto |\Omega|^{-M_q/2 + i}$~\cite{pichlerQuantum2015}. In the
above equation, of particular interest is the $N_m = 1$, $M_q = N$ case,
since that corresponds to the maximal genuine entanglement (across all bipartite cuts of qubits). We also consider $|\Omega| \to \infty$, since 
it is the most relevant for
metrology~\cite{gutierrez2023dissipative,groszkowski2022reservoir}.
Hence, we shall focus on obtaining the state
\begin{equation}
  \label{eqn: steady state ket}
  \ket{\Phi} \propto \sum \ket{S}_{i_1 i_2} \ket{S}_{i_3 i_4} \dots
  \ket{S}_{i_{N-1} i_N}
\end{equation}
where the summation in \eqref{eqn: steady state ket} runs over different
pairings of qubits $\{(i_1,i_2), (i_3,i_4), \dots (i_{N-1}, i_N)\}$ where
$i_{j} < i_{j+1}$. 
By a suitable detuning pattern, it is also possible to obtain the special
case where there is only one term in the sum, such that the system forms
dimerised pairs of qubits in the steady state. However, we will now
show that such schemes require a prohibitively long time to generate
high-fidelity, many-body entanglement.

\paragraph{Divergent timescale of preparing entangled dark states.---}
As mentioned in~\cite{pichlerQuantum2015}, the timescale required to form one dimer pair from $N=2$ qubits diverges as the target fidelity approaches one. This can also be seen by analysing the Liouvillian gap~\cite{albertSymmetries2014, manzanoHarnessing2018} (see Supplemental
Material~\cite{referenceToSupplementalMaterial}), but is analytically challenging for large $N$. By using a recently developed general framework for analysing quantum speed limits in dissipative state preparation~\cite{liuInitial2023}, we derive a lower bound on the time $T$ required to generate the state in \eqref{eqn:
most general steady state from zoller} for any system size $N$ (see Supplementary Material~\cite{referenceToSupplementalMaterial} for a derivation)
\begin{equation}
  \label{eqn: TQSL}
  T \geq T_\text{QSL} \propto
  \prod_{q=1}^{N_m} |\Omega|^{M_q/2} = |\Omega|^{N/2} \sim \left(\frac{1}{1-F}\right)^{N/4}.
\end{equation}
The preparation time diverges as $ |\Omega| \to \infty$, or equivalently as the fidelity $F$ to the target state in~\eqref{eqn: steady state ket} approaches unity. Crucially, for any fixed target fidelity $F$, the preparation time scales exponentially with the number of qubits $N$.

In the presence of \textit{any} spontaneous decay rate $\Gamma_f$
outside of the 1D bath, the time-independent scheme would fail when the
preparation time required exceeds $\sim 1/\Gamma_f$. From \eqref{eqn: TQSL},
we can estimate that the time-independent scheme fails for $N \gtrsim
\log(\Gamma/\Gamma_f)$, where $\Gamma = \gamma_L + \gamma_R$ is the total decay rate into the 1D
bath. This can be interpreted as a fundamental trade-off between fidelity
and speed, and highlights a severe limitation to the scalability of
such schemes. We now propose an exponentially faster scheme that circumvents
all these problems while retaining the robustness from dissipative
stabilization.

\paragraph{Exponentially faster scheme for many-body entanglement generation. ---}
Our scheme deviates from the previously proposed time-independent schemes in
two important aspects. Firstly, instead of a time-independent homogeneous
drive $\Omega_j = \Omega$, we consider $\Omega_j = \Omega(t)$ such that
$\Omega(0) = 0$ and $\Omega(t)$ is any non-decreasing real-valued
function of $t$. Secondly, all the detunings $\delta_j$ are zero, even
at zero chirality. In this case, with $\phi_{j k} \text{ mod }2\pi =
0$, in the master equation \eqref{eqn: main master eqn}, we have $\gamma_L
\mathcal{D}[c_L] + \gamma_R \mathcal{D}[c_R] = \Gamma \mathcal{D}[c]\rho$
where $c = \sum_{j=1}^N \sigma_j$, $\Gamma = \gamma_L + \gamma_R$,
and $(H_C)_{j k} = (i\Delta\gamma/2)(\sigma^\dagger_j
\sigma_k - \sigma_j \sigma^\dagger_k)$ where $\Delta \gamma =
\gamma_R - \gamma_L$.
We define the total coherent interaction term as $H(t) \equiv H_C +
H_\text{drive}(t)$, where $H_C = \sum_{j < k} (H_C)_{j k}$. 

Our scheme begins by choosing a target state $\ket{\Phi}$ of the form in
\eqref{eqn: steady state ket}, where in the summation, we have the freedom to
choose which different pairings of qubits to sum over. Let $\theta(\Omega(t))$ be a function where $\theta(\Omega(t) = 0) = 0$, $\theta(\Omega(t) = \infty) = \pi/2$. For example, $\theta(\Omega(t))$ could be
\begin{equation}
  \label{eqn: function for theta}
  \theta(\Omega(t)) = \frac{\pi}{2}(1- e^{-k\Omega(t)/\Gamma}), \,\, k>0,
\end{equation}
though many other examples exist. The main idea is that both the initial
state $\ket{g}^{\otimes N}$ and the target state $\ket{\Phi}$ at $\Omega(t)
\to \infty$ are instantaneous steady states, which means that if we can
generate the unitary evolution $U(\theta(\Omega))\ket{g}^{\otimes N} =
\cos(\theta(\Omega))\ket{g}^{\otimes N} - i \sin(\theta(\Omega))\ket{\Phi}$, then $\Omega \to \infty$ gives us $U(\theta(\Omega))\ket{g}^{\otimes N} = \ket{\Phi}$.
In practice, we do not require $\Omega(t)
\to \infty$, since at large $\Omega(t)$ such that $\theta(\Omega) =\pi/2-\epsilon$, 
$U(\theta)$ already prepares a state
$\ket{\psi(\theta)} \equiv U(\theta)\ket{g\dots g}$ with a fidelity of
$F = |\braket{\psi(\theta)}{\Phi}|^2 = \cos^2(\epsilon) \approx 1-\epsilon^2$ to
$\ket{\Phi}$. Hence, by a judicious choice of $\theta(\Omega(t))$ and
$\Omega(t)$, we can achieve a state $\ket{\psi(\theta)}$ that has very high
fidelity to $\ket{\Phi}$ at times much shorter than the dissipation timescale
$\Gamma^{-1}$. Using \eqref{eqn: function for theta} as an example, for $k \Omega(t)/\Gamma \approx 4$, we have $F \approx 0.999$. After preparing $\ket{\psi(\theta)}$ at a short time $t_f$, we keep $\Omega(t > t_f)$ constant. This causes the state to relax towards the steady state close to $\ket{\Phi}$.
Thus, our scheme works with a high fidelity even for a finite $\Omega$,
rendering its practicality. In short, our scheme moves along a trajectory
within the decoherence-free subspace spanned by $\ket{g}^{\otimes N}$ and
$\ket{\Phi}$ and is thus dissipation-stabilized.

To construct $U(\theta)$, we first define $X \equiv \ketbra{g^{\otimes
N}}{\Phi} + \ketbra{\Phi}{g^{\otimes N}}$ and then see that $U(\theta) =
\exp\left(-i \int_0^t (\partial_{t^\prime}\theta) X dt'\right)$
which means that the desired $U(\theta)$ can be generated by the
Hamiltonian $ H_u(t) = (\partial_{t}\theta) X$. Thus, we simply need to add
an extra time-dependent control field $H_\text{extra}(t) \approx H_u(t)$
to our system Hamiltonian $H(t)$. This extra time-dependent control field
would only need to be switched on from $t = 0$ to $t = t_f$ for some finite $t_f$ to generate $U(\theta)$, after which the time-dependence can be switched off and $\Omega(t)$ held constant. 
One might be concerned about spurious effects from the coherent interactions mediated by the 1D bath. While this can be entirely mitigated in $H_{\text{extra}}(t)$, we find that it is unnecessary.
The validity of the approximation $H_\text{extra}(t) \approx H_u(t)$ is
discussed in detail in the Supplemental
Material~\cite{referenceToSupplementalMaterial}, but here we note the
following two points. Firstly, the approximation is better for a smaller $\Delta \gamma$, with the best case being zero chirality ($\Delta
\gamma = 0$). This is actually an advantage when compared
to~\cite{pichlerQuantum2015} which requires $\Delta \gamma \neq 0$ when all
the detunings $\delta_{i}$ are zero. Secondly, by choosing $\partial_t\theta$
to be as large as possible, we can perform the transformation
$\ket{g}^{\otimes N} \to \ket{\psi(\theta)} \approx \ket{\Phi}$ in this decoherence-free subspace
arbitrarily quickly, which also improves the approximation $H_\text{extra}(t) \approx H_u(t)$.

We stress that while this protocol looks similar to the idea of counterdiabatic
driving in decoherence-free
subspaces~\cite{berry2009transitionless,Vacanti_2014,wu2017adiabatic} due to
the presence of an additional time-dependent control Hamiltonian, it is
different in many ways. Unlike counterdiabatic driving, the state
$\ket{\psi(\theta)}$ does not need to be an
instantaneous eigenstate of $H(t)$. In
fact, moving along the adiabatic trajectory in the Hilbert space as proposed
in~\cite{pichlerQuantum2015} requires $\Delta \gamma \neq 0$, whereas our
scheme allows for $\Delta \gamma = 0$. Thus, our scheme is fundamentally
different from the various shortcut-to-adiabaticity
schemes~\cite{odelin2019shortcuts}. In our computation of the extra driving
field $H_\text{extra}(t)$, unlike the various counterdiabatic driving
schemes, we do not require all the instantaneous eigenstates of $H(t)$. This
is highly advantageous in many situations where an exact
diagonalization of $H(t)$ is difficult, such as for large $N$. More details about the differences
between our proposed scheme and counterdiabatic driving can be found in the
Supplemental Material~\cite{referenceToSupplementalMaterial}.

In our scheme, the key part is implementing the
$X$ operator, which can be experimentally difficult for certain target states $\ket{\Phi}$ due to the many-body interactions required to
generate $X$. An example for $N=6$ qubits is shown in 
the Supplemental Material~\cite{referenceToSupplementalMaterial}.
However, when $\ket{\Phi}$ describes the state
of $N/2$ dimerised pairs, applying the above formalism gives us 
$U(\theta) = U_{i_1 i_2}(\theta)U_{i_3 i_4}(\theta)\dots
U_{i_{N-1} i_N}(\theta)$ where $U_{i_k i_{k+1}}(\theta) = \cos(\theta)
\mathbbm{1} - i \sin(\theta) X_{i_k i_{k+1}}$, and $X_{i_k i_{k+1}} =
\ketbra{gg}{S}_{i_k i_{k+1}} + \ketbra{S}{gg}_{i_k i_{k+1}}$ is a two-body
interaction term between qubits $i_k$ and $i_{k+1}$. 
$U(\theta)$ can then be generated by the Hamiltonian $H_u(t) =
(\partial_t \theta)X$ where $X = X_{i_1 i_{2}} + X_{i_2 i_{3}} + \dots +
X_{i_{N-1} i_{N}}$. Finally, we have $H_\text{extra}(t) \approx
H_u(t)$, which means that it suffices for the engineered control Hamiltonian to be $2$-qubit interactions. Explicitly, for geometrically
local dimer pairs, we have
$H_\text{extra}(t) \approx \sum_{k \, \text{odd}} V_{k,k+1}$ where $V_{k,k+1} = (\partial_t\theta) \left(\frac{1}{2}(\sigma_k^x - \sigma_{k+1}^x)
+ \frac{1}{2}(\sigma_{k}^x \sigma_{k+1}^z -
\sigma_k^z\sigma_{k+1}^x)\right)$.
Since the control Hamiltonian is local and can be applied in parallel, our protocol time is independent of $N$, which is exponentially faster than state-of-the-art time-independent schemes~\cite{pichlerQuantum2015,gutierrez2023dissipative} while still benefiting from dissipative stabilization. 

\begin{figure}[h]
  \centering
  \includegraphics[width=0.50\textwidth, trim={0cm 0cm 0cm 0cm},clip]{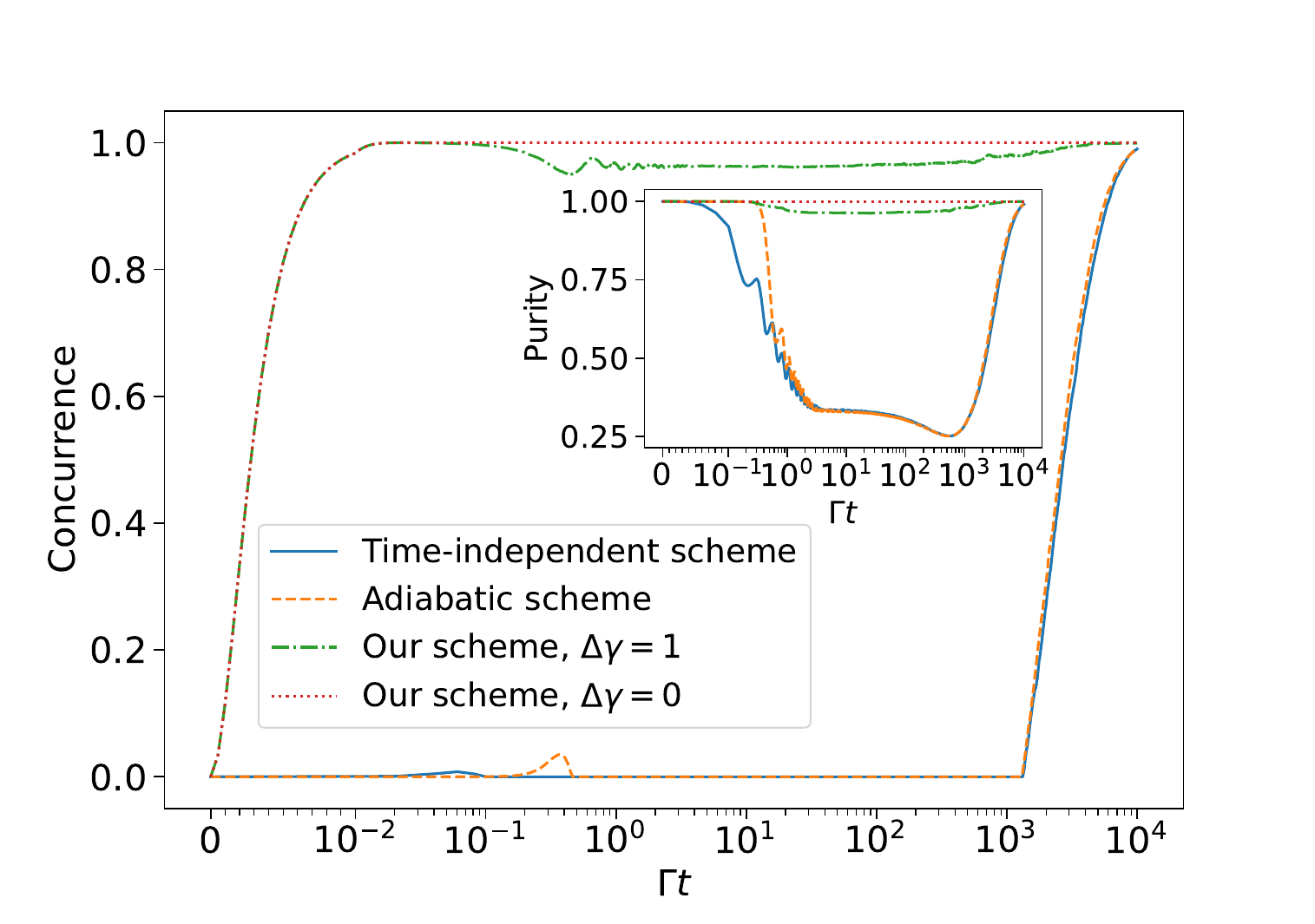}
  \caption[]{The case of $N = 8$ qubits forming
  $N/2 = 4$ geometrically local dimer pairs $\{(1,2), (3,4), (5,6), (7,8)\}$.
  Since all dimer pairs are treated equally, we plot the concurrence~\cite{wootters2001entanglement} and the purity (in the inset)
  between the qubits $(1,2)$ for various entanglement generation schemes
  mentioned in the main text. For our scheme, we use \eqref{eqn: function for
  theta} with $k = 10$. For both our scheme and the adiabatic scheme, we use the linear ramp function
  $\Omega(t)/\Gamma = m \Gamma t \Theta(t_f - t) + m\Gamma t_f \Theta(t - t_f)$ saturating at $t=t_f$, where
  $\Theta(t)$ is the Heaviside step function with $\Theta(0) = 1/2$, and with
  $m = 25$, $t_f = \Gamma^{-1}$, whereas for the time-independent scheme, we have
  $\Omega/\Gamma = 25$. For both the adiabatic scheme and the
  time-independent scheme, we have $\Delta \gamma/\Gamma = 1$, and also the
  appropriate detuning conditions as proposed in~\cite{pichlerQuantum2015}.
  The adiabatic schemes and the time-independent schemes are very similar
  after $t > t_f$ because the driving strengths $\Omega(t)$ become fixed
  after $t > t_f$. Clearly, only our scheme succeeds at short timescales
  $\Gamma t \ll 1$. }
  \label{fig: all schemes comparison N=8}
\end{figure}

\figref{fig: all schemes comparison N=8} shows the results of numerical
experiments comparing our scheme against previous proposals.
We also benchmark our
scheme against an adiabatic scheme.
As can be seen, at short timescales $\Gamma t \ll 1$, our
scheme achieves concurrence $\approx 1$ for the case where $\Delta \gamma =
0$ and concurrence $\approx 0.97$ for the case where $\Delta \gamma/\Gamma =
1$. 
On the other hand, the adiabatic scheme fails at timescales $\Gamma t \ll
1$ as the driving strength is modulated too quickly, violating the adiabatic
condition for open quantum systems~\cite{campo2016adiabaticity}. This is
corroborated by a sharp drop in purity between $0<t<t_f$. After $t > t_f$
where the driving strengths become fixed, the adiabatic and the
time-independent schemes become very similar. Our scheme is scalable and can be used to generate many dimer pairs simultaneously. 

\begin{figure}
  \centering
  \includegraphics[width=0.50\textwidth, trim={0cm 0cm 0cm 0cm},clip]{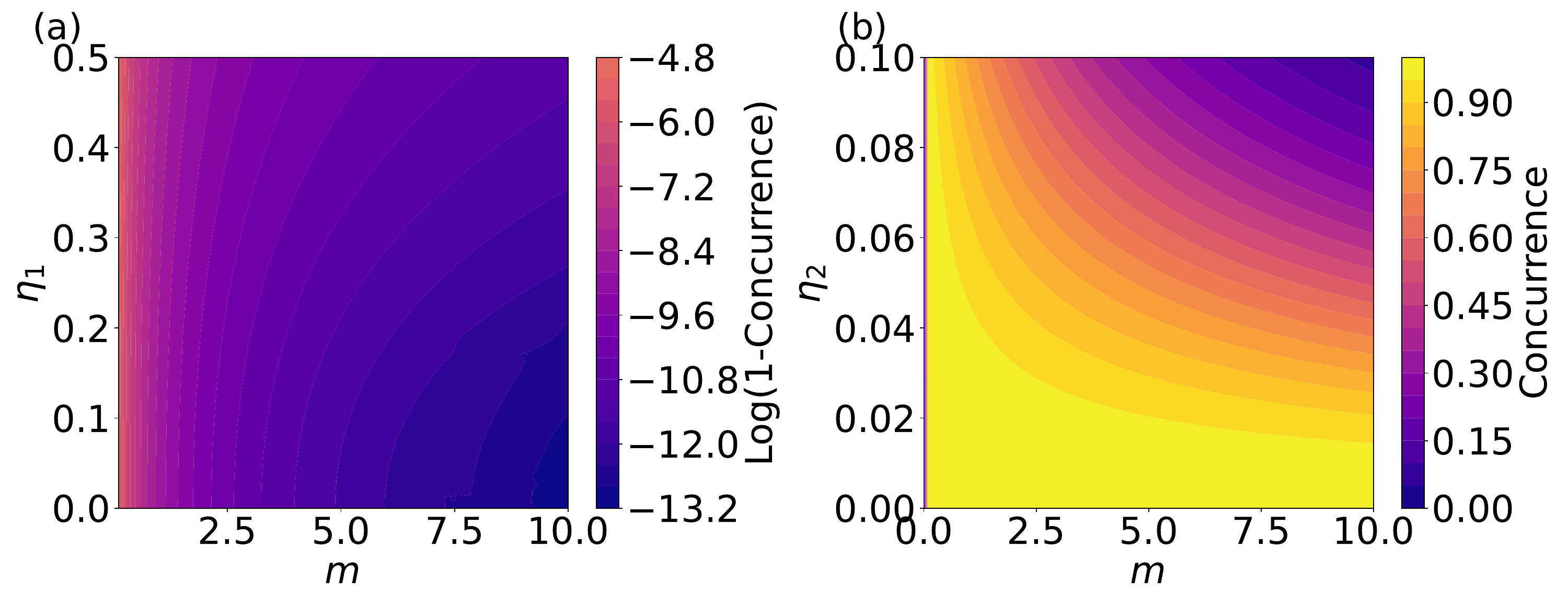}
  \caption[]{Analysis of the robustness of our scheme against noise. Here, we
  use \eqref{eqn: function for theta} for $\theta(\Omega(t))$ with $k = 10$
  and $\Omega(t) = mt$, and we consider the case where $\Delta \gamma = 0$ in
  generating $4$ dimerised pairs from $N=8$ qubits. Since the concurrences of
  all the dimers are the same, we use the concurrence of a dimer pair at the steady state to characterise
  the entanglement generated. In (a), since the values of concurrence $C$ of
  the final state obtained are all close to $1$, we plot $\text{log}(1-C)$
  against $\eta_1$ and $m$ while assuming $\eta_2 = 0$, and in (b), we plot
  the concurrence of the final state obtained as a function of $\eta_2$ and
  $m$ but with assuming $\eta_1 = 0$. From (a), since the values of the
  concurrence $C$ are all close to $1$, we see that our scheme is
  relatively insensitive to fluctuations in the driving strength $\Omega(t)$
  regardless of how fast we increase the driving, though there is still some
  trade-off. From (b), we see that there is a trade-off between the amount of
  noise allowed and the rate $m$ at which we can increase the driving
  strength $\Omega$.}
  \label{fig: robustness analysis}
\end{figure}

\paragraph{Robustness analysis. ---} 
We consider the robustness of our scheme to two types of noise which arise
from imperfect control. Let $\xi_1(t)$ and $\xi_2(t)$ be two independent
Gaussian white noise random variables with zero mean and unit variance. A
stochastic fluctuation in $H_\text{drive}(t)$ can be modelled by making the
replacement $H_\text{drive}(t) \to (1+\eta_1 \xi_1(t)) H_{\text{drive}}(t)$.
Similarly, a stochastic fluctuation in $H_\text{extra}(t)$ can be modelled by
making the replacement $H_\text{extra}(t) \to (1+\eta_2 \xi_2(t))
H_{\text{extra}}(t) $. Following~\cite{Ruschhaupt_2012,odelin2019shortcuts},
we average over the white noise random variables using Novikov's theorem for
white noise~\cite{novikov1965functionals} to obtain the following modified
master equation $\dot{\rho} = -i [H_\text{drive}(t) + H_{\text{extra}}(t), \rho] +\Gamma \mathcal{D}[c]\rho + \eta_1^2 \mathcal{D}[H_\text{drive}(t)]\rho + \eta_2^2 \mathcal{D}[H_\text{extra}(t)]\rho$.
Using $\theta(\Omega(t))$ from \eqref{eqn: function for theta} with
$k = 10$, and $\Omega(t) = mt , m>0$, we numerically study the effect of $\eta_i$
separately 
in \figref{fig: robustness analysis} for the $\Delta \gamma = 0$ case. Our scheme is robust against noise in
$H_\text{drive}$ regardless of how fast $\Omega(t)$ is increased. The reason is that our scheme works as long as $\Omega(t)/\Gamma \gg 1$ at large $t$, such that the fluctuations $\Omega(t)/\Gamma$ are insignificant. On the other hand, when dealing with noise in $H_\text{extra}(t)$,
there is a tradeoff between the amount of noise present $\eta_2$ and the
maximum $m$ allowed such that the concurrence remains high, which can be
explained by the adiabatic theorem for open quantum
systems~\cite{campo2016adiabaticity}. 

Another common source of noise is spontaneous decay outside of the 1D bath. As discussed earlier, the time-independent schemes fail completely for $N \gtrsim \log(\Gamma/\Gamma_f)$ where $\Gamma_f$ is the spontaneous decay rate, due to the exponentially long timescales needed. In contrast, our scheme is able to generate quantum dimers with high concurrence for any $N$ on the relevant system timescale $\Gamma^{-1}$, as long as $\Gamma_f/\Gamma \ll 1$, which is achievable in current experiments (see Supplementary Materials~\cite{referenceToSupplementalMaterial} for more details).

\paragraph{Discussion. ---} 
We present a new scheme for rapid, high fidelity generation of many-body
entanglement for qubits coupled to a 1D bath, which is also robust to noise.
Our scheme is exponentially faster than previously proposed time-independent
schemes
in~\cite{pichlerQuantum2015,ramosNon2016,ramos2014quantum,gutierrez2023dissipative},
and does not require chirality or specific detuning patterns on the qubits,
which makes it more convenient for experimental implementation. Our scheme
avoids the usual drawbacks of dissipative state preparation in open systems
such as the use of time-dependent dissipators or potentially unphysical
dynamics~\cite{Vacanti_2014}. Remarkably, to generate geometrically local
dimer pairs, we only require 2-qubit control Hamiltonians
$H_{\text{extra}}(t)$, which can be experimentally implemented in
superconducting qubits~\cite{blais2007quantuminformation,
sheldon2016procedure, mitchell2021hardware, wang2019synthesis}. Non-local
interactions between the dimers are suppressed by destructive interference.

Furthermore, recent experiments using superconducting qubits work with
free space spontaneous emission and dephasing decay rates of $\Gamma_f/2\pi \approx
15\text{ kHz}$ and $K_\phi/2\pi \approx 100\text{ KHz}$~\cite{zanner2022coherent}. Considering a typical decay rate
of a single qubit into a waveguide $\Gamma/2\pi \approx 15\text{MHz}$,
from~\figref{fig: all schemes comparison N=8}, it is clear that our scheme is faster that the superconducting qubit decoherence times. Since the time-independent scheme has been recently demonstrated experimentally with superconducting qubits~\cite{shah2024stabilizing}, it is a promising platform to realize our exponentially faster protocol. As potential future work, it is worth exploring the possibility of approximating the many-body interaction
terms in our general scheme using local driving terms,
following the formalism developed in~\cite{sels2017minimizing,
polkovnikov2023COLD} for counterdiabatic driving.

\begin{acknowledgments}
The IHPC A*STAR Team acknowledges support from the National Research
Foundation Singapore (NRF2021-QEP2-02-P01), A*STAR Career Development Award (C210112010), and A*STAR (C230917003, C230917007). K.H.L is grateful to the
National Research Foundation and the Ministry of Education, Singapore for
financial support. The Institute for Quantum Information and Matter is an
NSF Physics Frontiers Center. We thank William Chen, Leong-Chuan Kwek, Parth Shah, Richard Tsai, Sai Vinjanampathy and Frank Yang for helpful discussions.
\end{acknowledgments}

\bibliographystyle{apsrev4-2}
\bibliography{Biblio}

\onecolumngrid
\clearpage
\appendix 

\section{Liouvillian gap analysis of timescale for $N=2$}
\label{appendix: N=2 gap analysis}

\begin{figure}[h]
  \centering
  \includegraphics[width=0.5\textwidth, trim={0cm 0cm 0cm 0cm},clip]{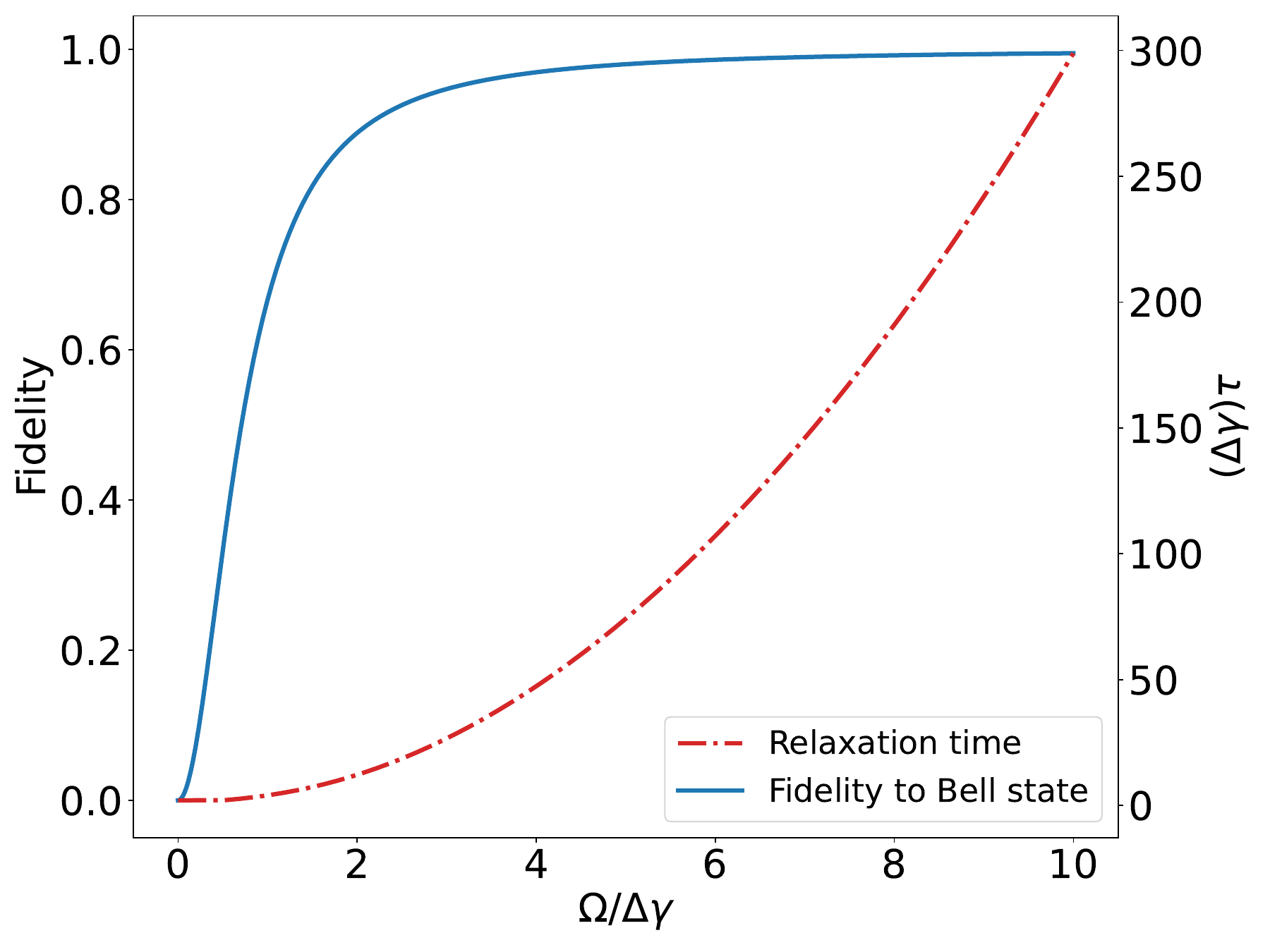}
  \caption[]{
  Time taken to reach the steady state (left axis) and the fidelity $F$ of the
  steady state to the dimer state (right axis) as a function of $\Omega/\Delta \gamma$. }
  \label{fig: fidelities and time taken as a function of driving}
\end{figure}

Here, we perform a Liouvillian gap analysis for the case where $N=2$ to
supplement the result in the main text, which uses a different
formalism~\cite{liuInitial2023} to prove that the timescale of forming a
dimer pair diverges. Here, we assume that the phase accumulated by the
excitation $\phi_{j k}$ as it travels between two system qubits is an
integer multiple of $2\pi$. In this case, the coherent interaction between the $i$th spin and the $j$th spin mediated by the 1D bath becomes $(H_C)_{j k} = (i\Delta\gamma/2)(\sigma^\dagger_j
\sigma_k - \sigma_j \sigma^\dagger_k)$ where $\Delta \gamma = \gamma_R -
\gamma_L$, and the collective decay into the 1D bath is given by $\Gamma
\mathcal{D}[c]\rho$, where $\Gamma = \gamma_L + \gamma_R$ and $c = \sum_{i=1}^N \sigma_i$. In this section, we consider the case where the
system qubits are driven using a time independent Hamiltonian
$H_\text{drive} = (\Omega/2)\sum_{i=1}^N \sigma_i^x$, though the driving Hamiltonian could also be time dependent.

To obtain the steady state for this case, we solve 
\begin{subequations}
  \label{appendix eqn: main master eqn}
  \begin{align}
    \dot{\rho} &= -i [H_S + \sum_{j<k} (H_C)_{j k}, \rho] +\Gamma \mathcal{D}[c]\rho \\
    H_S &= - \sum_{i=1}^N\delta_i \sigma_i^\dag \sigma_i + H_\text{drive}(t)
  \end{align}
\end{subequations}
\eqref{appendix eqn: main master eqn} for $\rho_{ss} = 0$ with $N = 2$. One way to do this
\cite{Gyamfi_2020} is to
vectorise the density matrix, i.e
\begin{subequations}
\begin{gather}
  \rho = \sum_{i,j}\rho_{ij}\ketbra{i}{j} \to |\rho) = \frac{1}{C}\sum_{i,j}\rho_{ij}\ket{i}\otimes \ket{j} \\
  A \rho_S \to (A\otimes \mathbbm{1}) |\rho_S) \\
  \rho_S B \to (\mathbbm{1} \otimes B^T) |\rho_S)
\end{gather}
\end{subequations}
where $|\rho)$ is the vectorised form of $\rho$, which itself can be thought of as a ket in the so-called Liouville space~\cite{Gyamfi_2020},
$A$ and $B$ are arbitrary operators, and $B^T$ denotes the transpose of
$B$. The constant $C$ in $|\rho)$ can be determined by normalising the state $|\rho)$  With the above description, the master equation becomes
\begin{equation}
  \frac{d
|\rho)}{dt} = L |\rho)  
\end{equation}
where
\begin{subequations}
  \begin{align}
    L =& -i(H\otimes \mathbbm{1} - \mathbbm{1}\otimes H^T) + \Gamma\left(c\otimes c^{*} - \frac{1}{2}(c^\dag c \otimes \mathbbm{1} - \mathbbm{1}\otimes c^\dag c)\right) \\
    H =& \frac{-i\Delta\gamma}{2} (\sigma^\dagger_2 \sigma_1 - \sigma_1 \sigma^\dagger_2) + \frac{\Omega}{2}(\sigma_1^x + \sigma_2^x)\\
    c =& \sigma_1 + \sigma_2.
  \end{align}
\end{subequations}
We note that for $N = 2$, we can get a unique steady state even without imposing extra conditions on $\delta_i$ \cite{pichlerQuantum2015}, and hence we have set all the $\delta_i = 0$.
Then, solving for the steady state $\rho_{ss}$ just reduces to finding the
nullspace of the matrix $L$. In this case, we have a unique steady state
\begin{subequations}
\label{appendix eqn: d=0 steady state}
\begin{align}
  \rho_{ss} &= \ketbra{s}{s} \\
  \ket{s} &= \frac{1}{\sqrt{2+(\frac{\Delta \gamma}{\Omega})^2}} \left(\frac{i\Delta \gamma}{\Omega} \ket{gg} - \ket{ge} + \ket{eg}\right).
\end{align}
\end{subequations}
Clearly, in the $\Omega/\Delta\gamma \to \infty$ limit, we obtain the dimer 
state $\ket{S} = (\ket{ge} - \ket{eg})/\sqrt{2}$ as the steady state of our
system with fidelity $1$.

From $L$, we can also calculate the (slowest)
timescale for the system to relax to the
steady state by calculating the inverse of the Liouvillian gap, which is the
largest non-zero real part of the eigenvalues of the matrix $L$
\cite{manzano2018Harnessing} (note that all the non-zero eigenvalues of $L$
have negative real parts).  For $\Omega \ll \Delta\gamma$, the Liouvillian gap is
$-\Delta\gamma/2$, which means that the system relaxes to the steady
state at a timescale $\tau = 2/\Delta\gamma$, independent of the driving
strength.  On the other hand, for $\Omega \gg \Delta\gamma$, the Liouvillian gap
is $-\Delta\gamma^3/(3\Omega^2)$, which means that the system relaxes to
the steady state at a timescale $\tau = 3\Omega^2/\Delta\gamma^3$. 
In
\figref{fig: fidelities and time taken as a function of driving}, we plot both the fidelity
of the steady state to the dimer state as well as the timescale $\tau$ (in
units of $\Delta\gamma^{-1}$) required to reach that steady state as a function
of $\Omega/\Delta\gamma$. In fact, we have 
\begin{equation}
  \label{appendix eqn: fidelity against time taken for steady state protocol}
  F = \frac{2\Delta\gamma \tau}{2\Delta\gamma \tau + 3}, \quad \Delta\gamma\tau = \frac{3}{2}\left(\frac{1}{1-F}-1\right)
\end{equation} 
where $F = \mel{S}{\rho_{ss}}{S}$ is the fidelity of the steady state to the
dimer state. This means that $\Delta\gamma \tau =
\mathcal{O}\left((1-F)^{-1}\right)$ as $F \to 1$. In other words, the time
required to obtain a dimer state as the steady state diverges with the
required fidelity of the state preparation procedure.  However,
repeating the Liouvillian gap analysis for $N > 2$ qubits quickly becomes
analytically intractable as $N$ increases, which is why in the main text we
used the formalism as described in \cite{liuInitial2023} to prove the
divergent timescale of obtaining the entangled steady state in general.

\clearpage
\section{Derivation of the divergent timescale of preparing entangled dark states}
\label{appendix: divegent timescale derivation}
For the system studied in the main text described by the master equation~\ref{appendix eqn: main master eqn}, it was shown~\cite{pichlerQuantum2015} that with
certain conditions on $\delta_i$ or in the chiral case $\gamma_L \neq
\gamma_R$,
it is possible to obtain the following multipartite entangled dark steady
state for even $N$
\begin{subequations}
  \label{appendix eqn: most general steady state from zoller}
  \begin{align}
    &\rho_{ss} = \ketbra{\Phi}{\Phi}, \quad \text{where } \ket{\Phi} =
    \prod_{q=1}^{N_m}\ket{M_q} \\
    \ket{M_q} =& a^{(0)}\ket{g}^{\otimes M_q} + \sum_{j_1 < j_2} a^{(1)}_{j_1, j_2}\ket{S}_{j_1 j_2} \ket{g}^{\otimes M_q -2} \nonumber \\
    \label{appendix eqn: multimer definition}
    &+ \dots + \sum a^{(M_q/2)}_{j_1,\dots j_{M_q}} \ket{S}_{j_1 j_2} \dots
    \ket{S}_{j_{M_q-1} j_{M_q}}.
  \end{align}
\end{subequations}
Here, we define $\ket{S}_{ij} =(\ket{e}_i\ket{g}_j - \ket{g}_j\ket{e}_i)/\sqrt{2}$ as a
singlet state (or a dimer pair) between qubits $i$ and $j$.
$\ket{\Phi}$ is a product of $N_m$ adjacent multimers $\ket{M_q}$, and each $\ket{M_q}$ is an entangled state over $M_q$ qubits as defined in \eqref{appendix eqn: multimer definition}, where $M_q$ is an even number.
Note that the summation in the last line of \eqref{appendix eqn: multimer definition} runs over all different pairings of
qubits $\{(j_1,j_2), \dots (j_{M_q-1}, j_{M_q})\}$ with $j_{k} < j_{k+1}$.
It can also be shown that  $a^{(i)} \propto |\Omega|^{-M_q/2 + i}$. 

Here, we want to use the general quantum speed limit framework for
dissipative state preparation introduced in \cite{liuInitial2023} to provide
a lower bound for the time $T$ required to generate the state in \eqref{appendix eqn:
most general steady state from zoller} for any even system size $N$. From \cite{liuInitial2023}, we have 
\begin{equation}
  T \geq T_\text{QSL} \propto \frac{1}{\mathcal{A}}
\end{equation}
where $\mathcal{A}$ in our case is simply $ |\Gamma| \times ||c^\dagger
\ketbra{\Phi}{\Phi}c||_F$ where $||X||_F \equiv \sqrt{\tr{X^\dagger X}}$ is
the Frobenius norm of the operator $X$. Here, we recall that $c = \sum_{i=1}^N
\sigma_i$. Now, from \eqref{appendix eqn: multimer definition}, we see that $c^\dagger
\ket{\Phi}$ annihilates all kets in the linear combination except the ket
$\ket{g}^{\otimes M_q}$. Thus, recalling that $a^{(0)} \propto
|\Omega|^{-M_q /2}$, we have $\mathcal{A} = |\Gamma| \prod_{q=1}^{N_m} |\Omega|^{-M_q/2}$, which gives us 
\begin{equation}
  \label{appendix eqn: TQSL}
  T \geq T_\text{QSL} \propto
  \prod_{q=1}^{N_m} |\Omega|^{M_q/2} = |\Omega|^{N/2} \sim \left(\frac{1}{1-F}\right)^{N/4},
\end{equation}
which diverges as $ |\Omega| \to \infty$, or equivalently as the fidelity $F$ to the target state in~\eqref{appendix eqn: steady state ket} approaches unity. Crucially, for any fixed target fidelity $F$, the preparation time scales exponentially with the number of qubits $N$. 

\clearpage  
\section{Comparison between our scheme and counterdiabatic driving}
\label{appendix: comparing our scheme to counterdiabatic driving}
In the counterdiabatic driving
scheme~\cite{berry2009transitionless,Vacanti_2014}, one often implements an
extra time-dependent Hamiltonian $H_\text{tqd}(t)$ to speed up the adiabatic
evolution due to a time-dependent Hamiltonian $H(t)$. Here, $H_\text{tqd}$
cancels out the term in $H(t)$ that leads to transitions between different
instantaneous eigenstates, and hence the system stays in its instantaneous
eigenstate at all times regardless of how large $\partial_t H(t)$ is. 
Determining the form of $H_\text{tqd}(t)$ is generally a difficult process,
as one needs to know all the time-dependent eigenstates of $H(t)$.
Furthermore, in open quantum systems, the concept of transitionless driving
is also further complicated by the need to maintain that the open systems evolution is a completely positive, trace-preserving (CPTP) map between density matrices at different times which might require one to engineer time-dependent
dissipators~\cite{Vacanti_2014}.

However, for the problem we are considering in our paper, in the case where
$\Delta \gamma \neq 0$, it is actually possible to use the idea of
counterdiabatic driving in decoherence free subspaces~\cite{wu2017adiabatic}.
We shall illustrate what we mean with the $N=2$ example. From \eqref{appendix eqn: d=0
steady state}, for the case where $\Delta \gamma \neq 0$, we see that by
assuming $\Omega(t)$ is a monotonically increasing function of $t$ such that
$\Omega(0) = 0$, then as we slowly increase $t$ from $0$ to $\infty$, we move
from the instantaneous eigenstate of $H(0)$ which is $\ket{gg}$ to the
instantaneous eigenstate of $H(\infty)$ which is $\ket{S}$. This follows from
the adiabatic theorem of quantum mechanics~\cite{kato1950adiabatic}.
Furthermore, since the instantaneous eigenstate of $H(t)$ is annihilated by
$c$ for all times $t$, when we use the technique of counterdiabatic driving,
we avoid the complications that follow from attempting to do counterdiabatic
driving for open quantum systems and we just need to consider the unitary
evolution case, as mentioned in~\cite{wu2017adiabatic}. We note here that
this scheme requries chirality, since if $\Delta \gamma = 0$, then there
would not be an adiabatic trajectory that connects $\ket{gg}$ and $\ket{S}$.

Having explained how one might use counterdiabatic driving to speed up the
many-body entanglement generation as proposed
in~\cite{pichlerQuantum2015,ramosNon2016,ramos2014quantum}, we note that the
key difference between our scheme and counterdiabatic driving is that, for
all intermediate times $t$ between $t=0$ and $t \to \infty$, there is no need
for our system state to be an instantaneous eigenstate of $H(t)$. This is
reflected in how our scheme allows for arbitrary choices of the function
$\theta(\Omega(t))$ that fulfil $\theta(\Omega(0)) = 0$ and
$\theta(\Omega(t\to\infty)) = \pi/2$. One practical implication of that is
that unlike counterdiabatic driving, we do not require chirality.
Furthermore, since counterdiabatic driving prevents transitions between all
instantaneous eigenstates of $H(t)$, the construction of $H_\text{tqd}$ would
require knowledge of all of the eigenstates of $H(t)$ for all times $t$. On
the other hand, since our scheme is only interested in constructing a
trajectory between $\ket{g\dots g}$ and $\ket{S}_{i_1 i_2} \ket{S}_{i_3 i_4}
\dots \ket{S}_{i_{N-1} i_N}$, we do not need to know all of the instantaneous
eigenstates of $H(t)$. 

To give a concrete example, we perform a comparison between our scheme and
the counterdiabatic scheme for the $N=2$ case. For the counterdiabatic
driving scheme, we use $\gamma_R = 1, \gamma_L = 0$ which gives us $\Delta
\gamma = 1$ and $\Gamma = 1$, whereas for our scheme, we use $\gamma_R = 0.5,
\gamma_L = 0.5$ which gives us $\Delta \gamma = 0$ and $\Gamma = 1$. For our
scheme, we also use 
\begin{equation}
  \label{appendix eqn: function for theta}
  \theta(\Omega(t)) = \frac{\pi}{2}(1- e^{-k\Omega(t)/\Gamma}), \,\, k>0,
\end{equation}
\eqref{appendix eqn: function for theta} with $k = 10$ for
$\theta(\Omega(t))$. In both cases, we use $\Omega(t) = mt$ with the same
value of $m$, and after $\Gamma t =1$, we switch off the extra control field
and we stop increasing the driving strength. For this case, we have:
\begin{subequations}
\label{appendix eqn: explicit form for htqd}
\begin{align}
  H_{\text{tqd}}(t) =&
   \frac{1}{2}\frac{m}{1 + 2m^2 t^2}\left(\sigma_1^x -
  \sigma_2^x + \sigma_1^x\sigma_2^z - \sigma_1^z\sigma_2^x\right)
  +\frac{1}{2} \frac{2 m^2 t}{1 + 6 m^2 t^2 + 8m^4 t^4} \left(\sigma_1^x\sigma_2^y + \sigma_1^y\sigma_2^x\right) \nonumber \\ 
  &+ \frac{1}{2}\frac{m}{1 + 6 m^2 t^2 + 8m^4 t^4} \left(-\sigma_1^x +
  \sigma_2^x + \sigma_1^x\sigma_2^z - \sigma_1^z\sigma_2^x\right)
  \end{align}
\end{subequations}
whereas for our scheme, we have $H_\text{extra}(t)$ as given in
\begin{subequations}
\begin{gather}
  \label{appendix eqn: H_extra combined}
  H_\text{extra}(t) \approx \sum_{k \,  \text{odd}} V_{k,k+1} \\
  \label{appendix eqn: H_extra two-body term}
  V_{k,k+1} = 
  (\partial_t\theta) \left(\frac{1}{2}(\sigma_k^x -
  \sigma_{k+1}^x) + \frac{1}{2}(\sigma_{k}^x \sigma_{k+1}^z -
  \sigma_k^z\sigma_{k+1}^x)\right).
\end{gather}
\end{subequations}
\eqref{appendix eqn:
H_extra combined}. Notice that $H_\text{extra}(t)$ is simpler than
$H_\text{tqd}(t)$ to implement experimentally as it has lesser many-body
interaction terms.
\begin{figure}[h]
  \centering
  \includegraphics[width=0.9\textwidth, trim={0cm 0cm 0cm 0cm},clip]{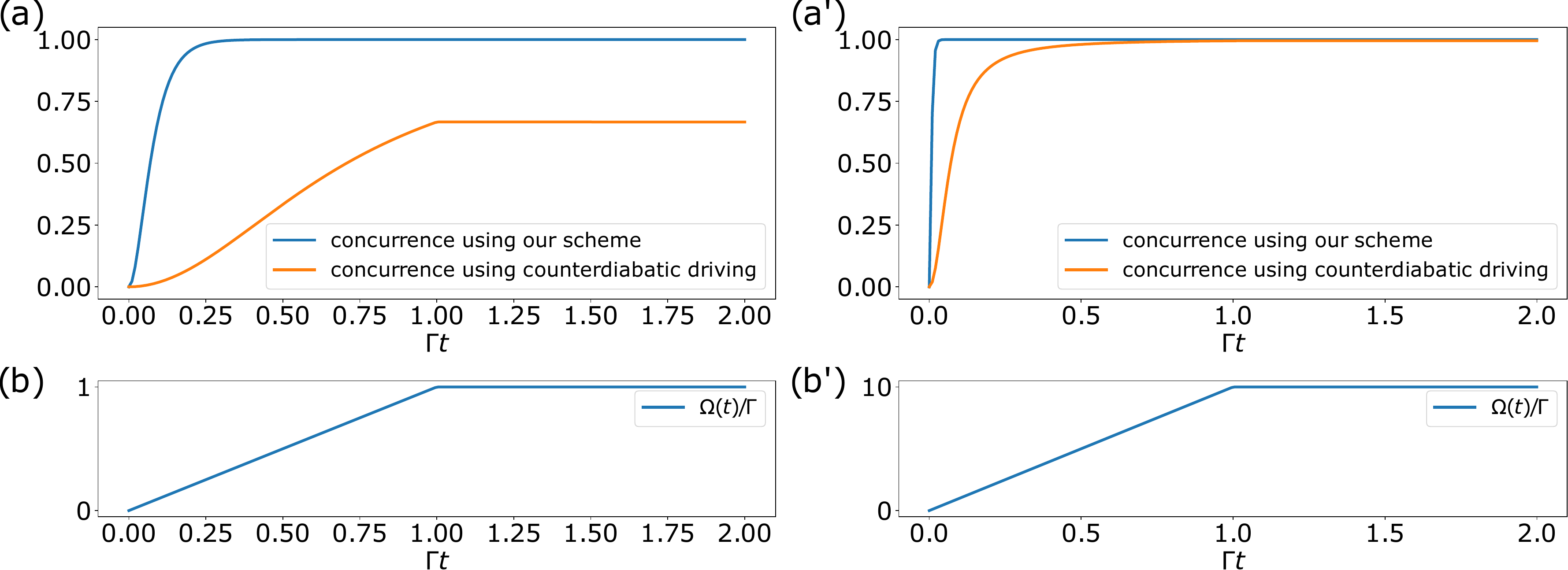}
  \caption[]{In (a) and (a'), for the $N = 2$ case, we plot the concurrence
  of our scheme and the counterdiabatic driving scheme for $m = 1$ and $m =
  10$ respectively, and in (b) and (b'), we show the time variation of the
  driving strengths $\Omega(t)$ against time. For our scheme, we use
  \eqref{appendix eqn: function for theta} with $k = 10$ for $\theta(\Omega(t))$.
  Clearly, our scheme outperforms the counterdiabatic driving scheme, since
  we do not need to stay in the instantaneous eigenstate of $H(t)$ given in
  \eqref{appendix eqn: d=0 steady state}.}
  \label{fig: comparison with TQD}
\end{figure}
The simulation results are given in \figref{fig: comparison with TQD}. Notice
that after $t = 1$, since the counterdiabatic driving case requires us to
stay in the instantaneous eigenstate of $H(t)$ given by \eqref{appendix eqn: d=0
steady state}, the largest concurrence we can get is
\begin{equation}
  C = \frac{2 m^2 t^2}{1+2m^2t^2}
\end{equation}
which is the fidelity of \eqref{appendix eqn: d=0 steady state} to the dimer state
$\ket{S}$. Hence as can be seen from \figref{fig: comparison with TQD}, for
the case where $m = 1$, we have a maximal concurrence of $2/3$ only. On the
other hand, for our scheme, we can very quickly get concurrence $1$ since we
do not need to follow the adiabatic trajectory to get the final state
$\ket{S}$.

\clearpage  
\section{Approximation of $H_\text{extra}$}
\label{appendix: approximating the extra driving field}
In writing $H_\text{extra}(t) = H_\text{u}(t) - H_C - H_d(t) \approx
H_u(t)$ in the main text, we made two approximations, first by ignoring
$-H_C$ and next by ignoring $-H_d(t)$ in $H_\text{extra}(t)$. Here, we study
the effect of both approximations. Here, we consider the problem of obtaining
the state
\begin{equation}
  \label{appendix eqn: steady state ket}
  \ket{\Phi} = \sum \ket{S}_{i_1 i_2} \ket{S}_{i_3 i_4} \dots
  \ket{S}_{i_{N-1} i_N}
\end{equation}
\eqref{appendix eqn: steady state ket} for general even $N$.

\subsection{Effect of ignoring $-H_C$}
Clearly, when $\Delta \gamma = 0$, ignoring $-H_C$ has no effect since $H_C =
0$. Hence, here we consider the case where $\Delta \gamma \neq 0$. For
general even $N$, the steady state of \eqref{appendix eqn: main master eqn} is
\eqref{appendix eqn: most general steady state from zoller}, where as mentioned in the
main text, we consider the case where we have only one multimer, i.e $N_m =
1$. For $\ket{M_q}$ in \eqref{appendix eqn: most general steady state from zoller}, 
the coefficient in front of the $\ket{g\dots g}$ term is proportional to
$\Delta \gamma^{N/2}$ for the case of zero detunings (i.e, $\delta_i = 0$ in \eqref{appendix eqn: main master eqn}).

Now, our scheme consists of switching on $H_\text{extra}(t)$ from $t = 0$ to
$t = t_f$ where $t_f \equiv \Omega^{-1}(\theta^{-1}(\pi/2 - \epsilon))$ is as
defined in the main text. After $t = t_f$, we switch off the extra driving
field and keep the driving strength $\Omega(t)$ at a constant finite value
$\Omega(t_f)$. As mentioned in the main text, this will give us a final state
$\ket{f}$ that is $1-\epsilon^2$ away in fidelity from \eqref{appendix eqn:
steady state ket}. At this point, since $\Delta \gamma \neq 0$ and since
$\Omega(t)$ is finite, the state $\ket{f}$ is not an instantaneous eigenstate
of $H(t)$. Hence, there will be transitions induced by $H(t)$ on $\ket{f}$ to
all the instantaneous eigenstates of $H(t)$, some of which are not dark
states. Hence, the state becomes mixed and the fidelity to \eqref{appendix eqn: steady
state ket} drops. Note that depending on the choice of $\theta(t)$, since
$|H_\text{extra}(t)| \propto \partial_t \theta$, if $\partial_t \theta
\approx 0$ before $t_f$, then the above effect becomes more pronounced since
we obtain the final state $\ket{f}$ before $t_f$. However, if
$\Delta\gamma/\Omega(t)$ is small enough, then the probability amplitude of
$\ket{g\dots g}$ component in the dark state given in \eqref{appendix eqn: most
general steady state from zoller} will be small, which means that the overlap
between the $\ket{f}$ and the dark state will be large. This means that the
transitions induced by $H(t)$ on $\ket{f}$ will largely be to the dark state,
which means that the fidelity to \eqref{appendix eqn: steady state ket} remains high. 

We illustrate the above with the $N=8$ case where we form $4$ dimers, i.e
where our target steady state is $\ket{\Phi} =
\ket{S}_{12}\ket{S}_{34}\ket{S}_{56}\ket{S}_{78}$. For our scheme, we use
$\gamma_R = 1, \gamma_L = 0$ which gives us $\Delta \gamma = 1$ and $\Gamma =
1$. Here, we consider $\Omega(t) = mt, m > 0$, and after $t = t_f =
1/\Gamma$, we switch off $H_\text{extra}(t)$ and fix $\Omega(t)$ at the
constant value $\Omega(t_f)$. We also choose $\theta(\Omega(t))$ according to
\eqref{appendix eqn: function for theta} with $k=10$. In \figref{fig: effect of
chirality on H_extra approximation} we show the extent of the negative effect
that chirality has on our system at different values of $\Omega(t_f)$. 

\begin{figure}
  \centering
  \includegraphics[width=0.9\textwidth, trim={0cm 0cm 0cm 0cm},clip]{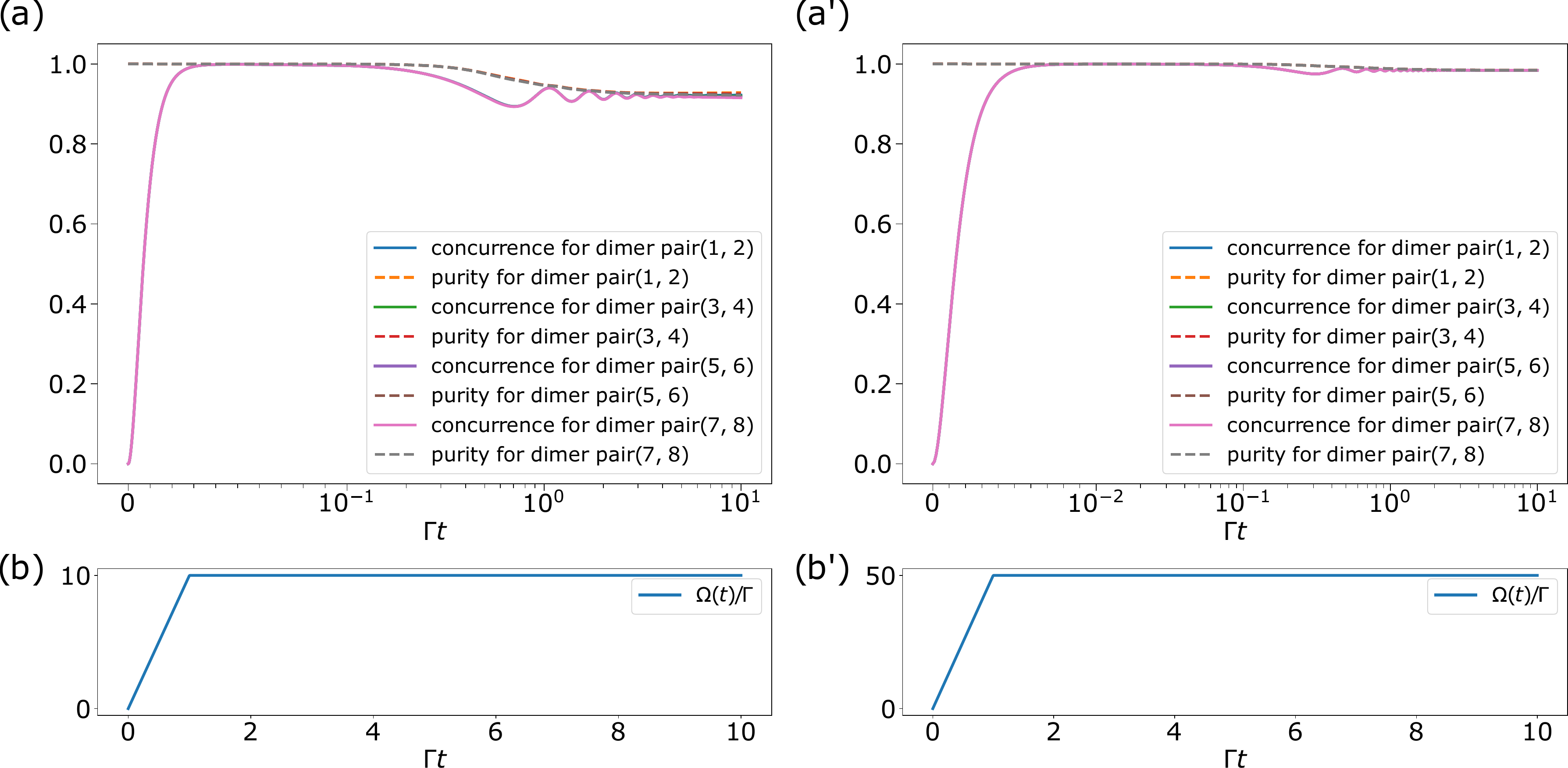}
  \caption[]{Here, we study the effect of ignoring $-H_C$ in $H_\text{extra}$
  when 
  $\Delta \gamma/\Gamma = 1$. In (a) and (a'), for the case where we have $N
  = 8$ qubits forming $4$ dimer pairs, we plot both the concurrences of the dimer
  pairs as well as the purity of the dimer pairs for $m = 10$ and $m = 50$
  respectively. In (b) and (b'), we plot the variation of the driving
  strength $\Omega(t)$ as a function of time.
  As can be seen, in both (a) and (a'), we quickly obtain the state
  $\ket{\Phi}$, even before we switch off $H_\text{extra}(t)$. This is
  because the form of \eqref{appendix eqn: function for theta} with $k=10$ causes
  $H_\text{extra}(t)$ to be very close to zero even before $\Gamma t = 1$.
  Then, in (a), since $\Omega$ is not high enough, the state $H(t)$ induces
  transitions on $\ket{\Phi}$ and together with the jump operators, we get a
  mixed state. In (a'), since $\Omega$ is high enough, the transitions
  induced by $H(t)$ are largely onto the dark state and hence the state
  largely remains pure.}
  \label{fig: effect of chirality on H_extra approximation}
\end{figure}

\subsection{Effect of ignoring $-H_\text{drive}(t)$}
Firstly, defining the triplet state $\ket{T}_{ij} \equiv (\ket{e}_i \ket{g}_j
+ \ket{g}_i\ket{e}_j)/\sqrt{2}$, the $H_\text{drive}(t)$ term can be written as
$H_\text{drive}(t) = H_\text{drive}^\prime(t) + O(t)$ where
$H_\text{drive}^\prime(t) = \frac{\Omega(t)}{\sqrt{2}}
\sum(\ketbra{T}{gg}_{i_1 i_2} + \dots + \ketbra{T}{gg}_{i_{N-1} i_N} +
\text{H.c})$ and $O(t) = \Omega(t)\sum(\ketbra{T}{ee}_{i_1 i_2} + \dots +
\ketbra{T}{ee}_{i_{N-1} i_N} + \text{H.c})/\sqrt{2}$, and where the sum is
over all pairs $\{(i_1,i_2), (i_3,i_4), \dots (i_{N-1}, i_N)\}$ with $i_{j} <
i_{j+1}$. In this decomposition, $O(t)$ annihilates the decoherence-free
subspace, and since it commutes with $H_u(t)$, we can
ignore the effect of $O(t)$. Hence, it remains to study the effect of ignoring $-H_\text{drive}^\prime(t)$. 

\begin{figure}
  \centering
  \includegraphics[width=0.9\textwidth, trim={0cm 0cm 0cm 0cm},clip]{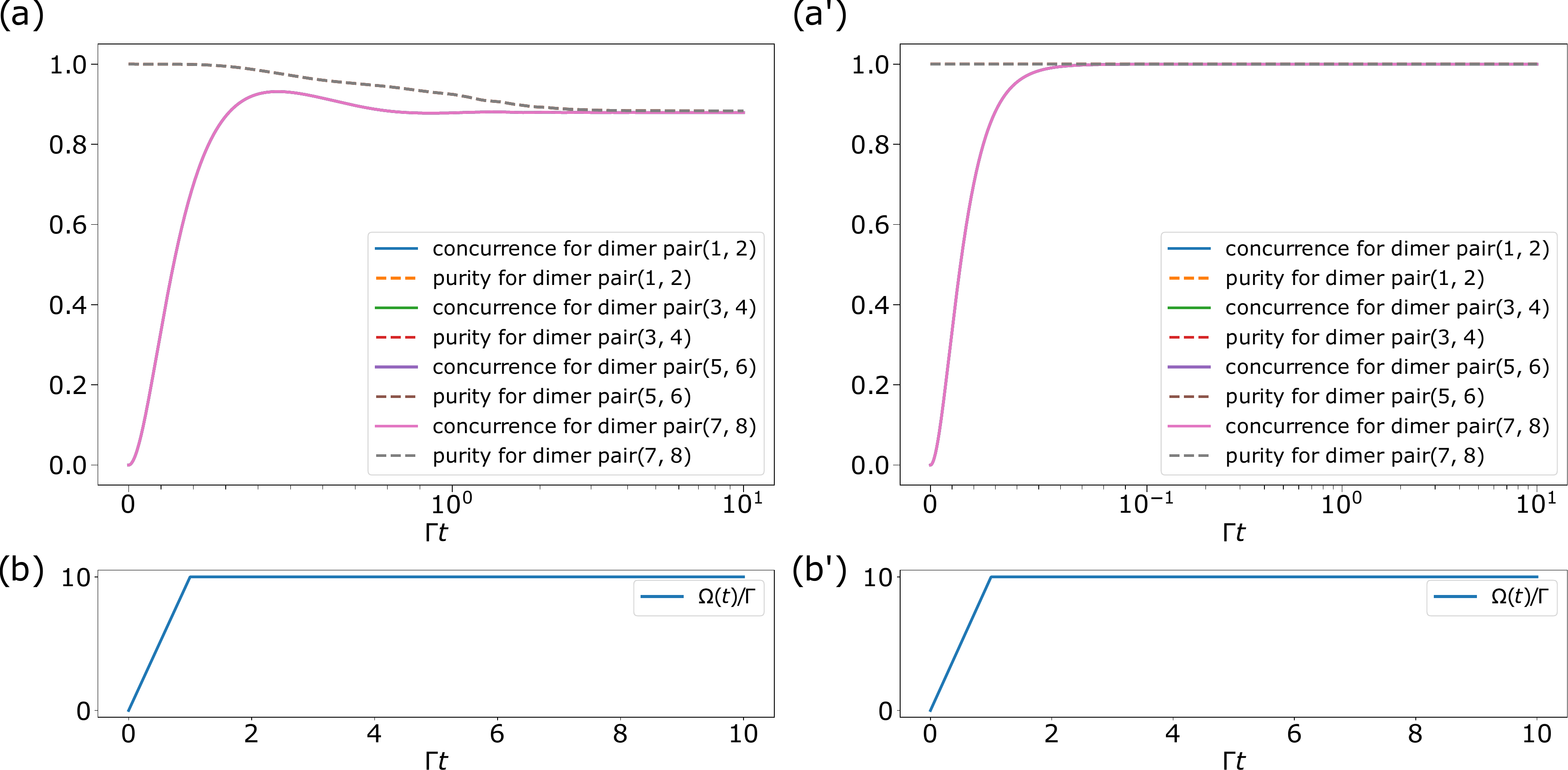}
  \caption[]{Here, we study the effect of ignoring $-H^\prime_\text{drive}$
  in $H_\text{extra}$ when $\Delta \gamma = 0$. In (a) and (a'), for the case
  where we have $N = 8$ qubits forming $4$ dimers, we plot both the
  concurrences of the dimer pairs as well as the purity of the dimer pairs
  for $m = 10$. In (b) and (b'), we plot the time variation of the driving
  strength $\Omega(t)$. In both (a) and (a'), we use \eqref{appendix eqn: function for
  theta} for $\theta(\Omega(t))$ with $k = 0.5$ and $k=5$ respectively.
  Clearly, it is permissible to ignore $-H^\prime_\text{drive}$ at large $k$.
  This is because the system goes from $\ket{g \dots g}$ so rapidly that
  $-H^\prime_\text{drive}$ has no effect.}
  \label{fig: effect of ignoring H_drive^prime}
\end{figure}

The idea is that if the transformation $\ket{g \dots g} \to \ket{\Phi}$ due
to $H_u(t)$ is much quicker than the transformation $\ket{g \dots g} \to
\ket{T}_{i_1 i_2}\dots\ket{T}_{i_{N-1} i_N}$ due to
$H_\text{drive}^\prime(t)$, then the effect of ignoring
$-H_\text{drive}^\prime(t)$ is negligible. This can be done in many ways, for
example by choosing $\theta(\Omega(t))$ according to \eqref{appendix eqn: function for
theta} with a large value of $k$. An example for the $N=8$ case where we form
$4$ dimers, i.e where our target steady state is $\ket{\Phi} =
\ket{S}_{12}\ket{S}_{34}\ket{S}_{56}\ket{S}_{78}$ is shown in \figref{fig:
effect of ignoring H_drive^prime} below. Since we studied the effect of a
non-zero $\Delta \gamma$ above, here we set $\Delta \gamma = 0$ to solely
study the effect of ignoring $H_\text{drive}^\prime(t)$. As can be seen from
\figref{fig: effect of ignoring H_drive^prime}, for small $k$ such that the
transformation $\ket{g \dots g} \to \ket{\Phi}$ is slow, the effect of
ignoring $H_\text{drive}^\prime(t)$ leads to quite substantial errors, but
for large $k$, we can safely ignore $-H_\text{drive}^\prime(t)$.

\clearpage  
\section{Effect of single spin spontaneous emission into free space on our
scheme}
\label{appendix: effect of single spin spontaneous emission}
Here, we consider the effect that single spin spontaneous emission on
entanglement generation for both our scheme and the time-independent
schemes in~\cite{pichlerQuantum2015,ramosNon2016,ramos2014quantum}. Single
spin spontaneous emission into free space can be modelled by adding an
additional term $\Gamma_f \sum_{i=1}^N\mathcal{D}[\sigma_i]\rho$ into
\eqref{appendix eqn: main master eqn}. The presence of $\Gamma_f$ means that the pure
state $\ket{\Phi}$ as defined in \eqref{appendix eqn: most general steady state from
zoller} is no longer a steady state of the master equation dynamics. This
leads to a reduction in the long-time purity as well as the amount of
long-time entanglement generated for the time-independent schemes, as
mentioned in~\cite{pichlerQuantum2015,ramos2014quantum}. Our scheme also
suffers a similar reduction in long-time purity as well as long-time
entanglement generated, as our scheme is an accelerated form of the
time-independent schemes.

However, we note that for waveguide QED systems, high $\beta$
factors~\cite{sheremetWaveguide2023} of up to $\beta = 0.99$ have been
experimentally demonstrated, where the $\beta$ factor is defined as $\beta =
\Gamma /(\Gamma + \Gamma_f)$, which is the ratio of the system radiative
decay into the $1$D bath over the total radiative decay rate of the system.
Thus, in terms of the timescale defined by $\Gamma^{-1}$, $\Gamma_f$ is
largely negligible as long as $\Gamma_f$ is small enough. Since our
accelerated scheme is able to generate entanglement at time $\Gamma t \leq
1$, the maximal entanglement generated by our scheme remains high.
Furthermore, the entanglement generated is still relatively long lived in
terms of the timescale defined by $\Gamma^{-1}$.
On the other hand, because the time-independent schemes require a
long time for entanglement generation, the presence of $\Gamma_f$ would
affect the maximal entanglement that is generated.

To see the effects mentioned above, we consider the case of preparing $N/2$ local dimers from $N$ qubits, i.e we want to prepare $\ket{\Phi}
= \ket{S}_{1,2}\ket{S}_{3,4}\dots\ket{S}_{N-1,N}$. For the time-independent
scheme, we consider the non-chiral case with the detuning pattern on the
qubits $[\delta_1, -\delta_1, \delta_2, -\delta_2, \dots \delta_{N/2},
-\delta_{N/2}]$ where $\delta_i \neq \delta_j$. This is the detuning pattern
that leads to the formation of local dimer pairs with the time-independent
scheme~\cite{pichlerQuantum2015}. We will use the same parameters as the
recent experimental work Ref.~\cite{shah2024stabilizing} where the authors
implemented the time-independent scheme in~\cite{pichlerQuantum2015}. This
means that we use $\Omega/\Gamma \approx 5$ for the time-independent scheme
with $\delta_i/\Omega \approx 1/2$ and $\Delta \gamma = 0$. On the other
hand, for our time-dependent scheme, we use the same value of
$\Omega/\Gamma$ but instead we use $\delta_i = 0$, since our
scheme does not need the detunings to produce the local dimers. We will
consider both $\Gamma_f/\Gamma = 0.1$ which is demonstrated
in~\cite{shah2024stabilizing} for implementing the time-independent scheme
and also $\Gamma_f/\Gamma = 0.01$ which is achievable with current
experimental techniques for superconducting qubit
platforms~\cite{zanner2022coherent,sheremetWaveguide2023}.

Firstly, we show that for the above experimental parameters, the steady state
concurrence $C$ of each dimer pair and its fidelity to the Bell state $F$ dips
significantly below $1$ even for small values of $\Gamma_f$ for $N \geq 4$.
This numerically demonstrates our claim above that in the steady state, both
the time-independent scheme and our scheme are adversely affected by
free-space decay. This is intuitively obvious from the fact that the steady
state is the $t\to \infty$ state of the master equation dynamics, which means
even a small amount of free-space decoherence $\Gamma_f$ can have a large
effect. A summary of results for $C$ and $F$ in the steady state is shown in
Table~\ref{table: steady state FC}. We note that though larger values of
$\delta_i$ can have a positive effect on entanglement generation (as measured by $C$), this effect becomes largely negligible as the number of system qubits $N$ increases.

\begin{table}[h]
\centering
\begin{tabular}{|c|cc|cc|}
\hline
\multirow{2}{*}{$N$, $\Gamma_f$} & \multicolumn{2}{c|}{$\delta_i \approx \Omega / 2$}       & \multicolumn{2}{c|}{$\delta_i \ll \Omega$}               \\ \cline{2-5} 
                                 & \multicolumn{1}{c|}{Steady state $C$} & Steady state $F$ & \multicolumn{1}{c|}{Steady state $C$} & Steady state $F$ \\ \hline
$N=4, \Gamma_f = 0.01, 0.10$     & \multicolumn{1}{c|}{$0.137, 0.083$}   & $0.684, 0.667$   & \multicolumn{1}{c|}{$0, 0$}           & $0.489, 0.486$   \\ \hline
$N=6, \Gamma_f = 0.01, 0.10$     & \multicolumn{1}{c|}{$0.015, 0$}       & $0.650, 0.641$   & \multicolumn{1}{c|}{$0, 0$}           & $0.489, 0.484$   \\ \hline
$N=8, \Gamma_f = 0.01, 0.10$     & \multicolumn{1}{c|}{$0, 0$}           & $0.631, 0.626$   & \multicolumn{1}{c|}{$0, 0$}           & $0.488, 0.480$   \\ \hline
\end{tabular}
\caption{Here we numerically solve for $\dot{\rho} = 0$ in \eqref{appendix eqn: main
master eqn} with an additional dissipative term $\Gamma_f
\sum_{i=1}^N\mathcal{D}[\sigma_i]\rho$ to find the steady state in the
presence of free space spontaneous emission with rate $\Gamma_f$. Parameters
used are: $\Delta \gamma = 0, \Omega/\Gamma = 5$, with the detuning pattern
on the qubits given by $[\delta_1, -\delta_1, \delta_2, -\delta_2, \dots
\delta_{N/2}, -\delta_{N/2}]$ where $\delta_i = \Omega/2 + 0.01(i-1)$ (in
units of $\Gamma$) for the $\delta_i \approx \Omega/2$ case. We also consider
the $\delta_i \ll \Omega$ case to study the effect that $\delta_i$ has on the
steady state concurrence. For the $\delta_i \ll \Omega$ case, we have
$\delta_i = 0.01i$ (in units of $\Gamma$). From the steady state obtained, we
calculate fidelity $F$ of the $i\text{th},i\text{th}+1$ qubit pair to the
Bell state, as well as the concurrence $C$ of the $i\text{th},i\text{th}+1$
spin pair. We then tabulate the average values of fidelity and concurrence
across all qubit pairs. Note that with $\Gamma_f = 0$, we would expect $F,C
\approx 1$ since the detuning pattern chosen as well as the high value of
$\Omega/\Gamma$ would lead to the formation of perfect dimerised
pairs~\cite{pichlerQuantum2015}.}
\label{table: steady state FC}
\end{table}

Secondly, we show that as mentioned above, the high $\beta$ factors in
current experiments mean that even though the steady state has a low
concurrence in the presence of $\Gamma_f$, we can still achieve reasonably
long-lived entanglement (in units of the relevant timescale $\Gamma^{-1}$) if
the entanglement can be generated fast enough. This would be possible with
our scheme, but not with the time-independent schemes which take
prohibitively long to generate the entanglement. More precisely, we note that
the time-independent scheme fails when the preparation time required
exceeds $\sim 1/\Gamma_f$, which means that from \eqref{appendix eqn: TQSL}, we can
estimate that the time-independent scheme to fail for $N \gtrsim
\log(\Gamma/\Gamma_f)$. Examples for $N = 4,6,8$ are shown in \figref{fig:
spontaneous emission main plot}. As can be seen, our scheme is able to
achieve a high maximal concurrence of $1$ for each dimer pair. Furthermore,
the entanglement generated by our scheme is relatively long-lived. This is
because our entanglement generation scheme is rapid and hence perfect
quantum dimer pairs are formed before the spontaneous emission has any
appreciable effect on our system. On the other hand, we see that the
time-independent scheme fails to generate any substantial level of
concurrence before the effect of spontaneous emission causes the concurrence
to drop back to near zero. We note here that as shown in Table~\ref{table:
steady state FC}, the large $\delta_i$ in the time-independent scheme allows
for a small non-zero concurrence in the steady state in the $N = 4$ case.
However, as can be seen, the effect of the large $\delta_i$ on the steady
state concurrence becomes negligible for $N \geq 6$. We show the maximal
concurrence generated by both our scheme and the time-independent scheme
in Table~\ref{table: maximal concurrence reached}.

\begin{figure}[h]
  \centering
  \includegraphics[width=0.95\textwidth, trim={0cm 0cm 0cm 0cm},clip]{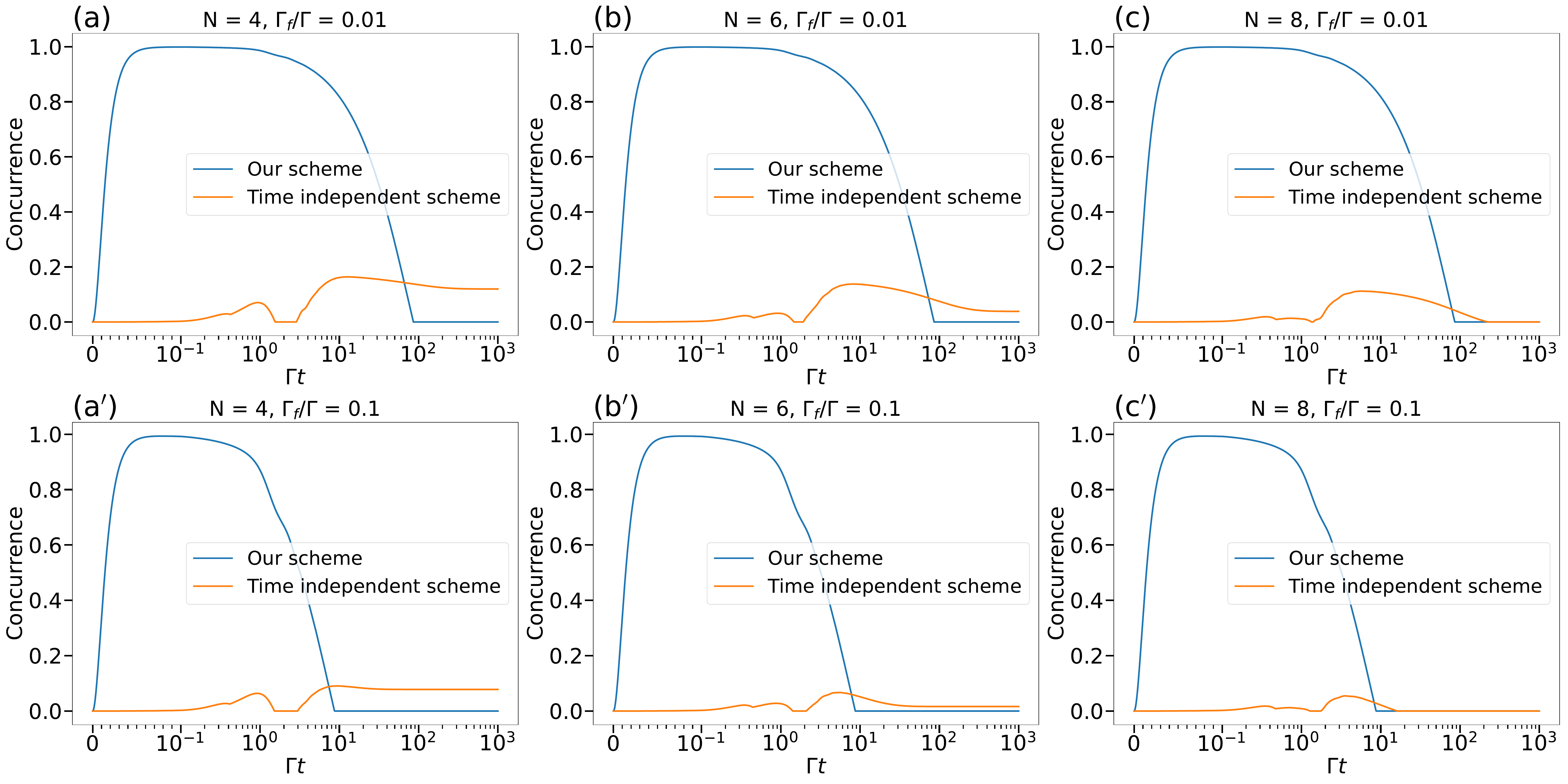}
  \caption[]{
  Numerical experiments for the concurrence against time for both the time-independent scheme and for our scheme.
  Here, we use the same
  experimental parameters as in Table~\ref{table: steady state FC}. Plots (a), (b), (c) are for $N = 4,6,8$ qubits respectively, with $\Gamma_f/\Gamma = 0.01$. Plots (a$^\prime$), (b$^\prime$), (c$^\prime$) are also for $N = 4,6,8$ qubits respectively, but with $\Gamma_f/\Gamma = 0.1$.
  At long times $\Gamma t$, the concurrence of both our scheme and the
  time-independent scheme drops to near-zero, in agreement with
  Table~\ref{table: steady state FC}. However, since our scheme is rapid,
  the entanglement is generated rapidly at $\Gamma t \ll 1$ before the
  effects of spontaneous emission kicks in, which allows our scheme to
  achieve a maximal concurrence of $1$. Furthermore, the entanglement
  generated by our scheme is relatively long-lived in units of
  $\Gamma^{-1}$, especially for smaller values of $\Gamma_f$. Thus, our
  scheme achieves a high maximal concurrence of $1$. 
  On the other hand, the time-independent scheme generates
  entanglement too slowly and the effect of spontaneous emission kicks in
  even before any appreciable entanglement is generated, as can be seen by the concurrence dipping to near-zero at long times.}
  \label{fig: spontaneous emission main plot}
\end{figure}

\begin{table}[h]
\centering
\begin{tabular}{|c|c|c|}
\hline
$N$, $\Gamma_f$              & \begin{tabular}[c]{@{}c@{}}Maximal concurrence \\ reached by our scheme\end{tabular} & \begin{tabular}[c]{@{}c@{}}Maximal concurrence reached \\ by the time-independent scheme\end{tabular} \\ \hline
$N=4, \Gamma_f = 0.01, 0.10$ & $1, 1$                                                                               & $0.163,  0.091$                                                                                       \\ \hline
$N=6, \Gamma_f = 0.01, 0.10$ & $1, 1$                                                                               & $0.137, 0.067$                                                                                        \\ \hline
$N=8, \Gamma_f = 0.01, 0.10$ & $1, 1$                                                                               & $0.112,  0.055$                                                                                       \\ \hline
\end{tabular}
\caption{Maximal concurrence generated by both our scheme and by the time-independent scheme, using the same experimental parameters as per Table~\ref{table: steady state FC}. Clearly, only our scheme achieves a high maximal concurrence. }
\label{table: maximal concurrence reached}
\end{table}

{
\section{Effect of unequal local light-matter coupling}
The effect of unequal local light-matter coupling in our system is captured
by the term $\sum_{i=1}^N \delta_i \sigma_i^\dagger \sigma_i$ in \eqref{appendix eqn:
main master eqn} in the form of different detunings of the spins to the
driving field $H_\text{drive}$. To obtain dimerised pairs in the steady state
in the case of time-independent driving, it is required that the detunings
on the spins follow a specific pattern
\begin{equation}
  \label{appendix eqn: Zoller detuning pattern dimers}
  [\delta_1, -\delta_1, \delta_2, -\delta_2, \dots \delta_{N/2-1},
  -\delta_{N/2}]
\end{equation}
where $\delta_i \neq \delta_j$, for the case where $N >
2$~\cite{pichlerQuantum2015}. On the other hand, for our protocol with
time-dependent driving, we actually do not require engineering such a
detuning pattern on the qubits, i.e our protocol works even with $\delta_i =
0$ for all qubits. i
In this section, we study the effect of $\delta_i \neq 0$ for our protocol,
which may be a result of experimental imprecision in the creation of the
qubits.

When $\delta_i \neq 0$ but $\delta_i$ still obeys the detuning pattern as per
\eqref{appendix eqn: Zoller detuning pattern dimers}, our protocol would still produce
dimerised pairs in the steady state. In this case, the concurrence of the
dimerised pairs would depend on the ratio $\Omega(t_f)/\text{max}(\delta_i)$
where we recall that $t_f$ is the time after which the strength of the
driving field $\Omega(t)$ remains constant. This is because while the
detuning pattern in \eqref{appendix eqn: Zoller detuning pattern dimers} guarantees
the formation of dimerised pairs as per~\cite{pichlerQuantum2015}, the
singlet fraction of each dimer pair depends on the relative strength of
$\Omega$ as compared to the other parameters in the system. Since the
detuning pattern obeys \eqref{appendix eqn: Zoller detuning pattern dimers}, we note
that after a sufficiently long time (see Appendix~\ref{appendix: divegent
timescale derivation} for the lower bound on the time taken), the time
independent scheme would also produce the state of dimerised pairs. A
numerical example comparing our scheme to the time-independent scheme in this
case is shown in Figure~\ref{fig: effect of detunings}a.
\begin{figure}[h]
  \centering
  \includegraphics[width=0.9\textwidth, trim={0cm 0cm 0cm 0cm},clip]{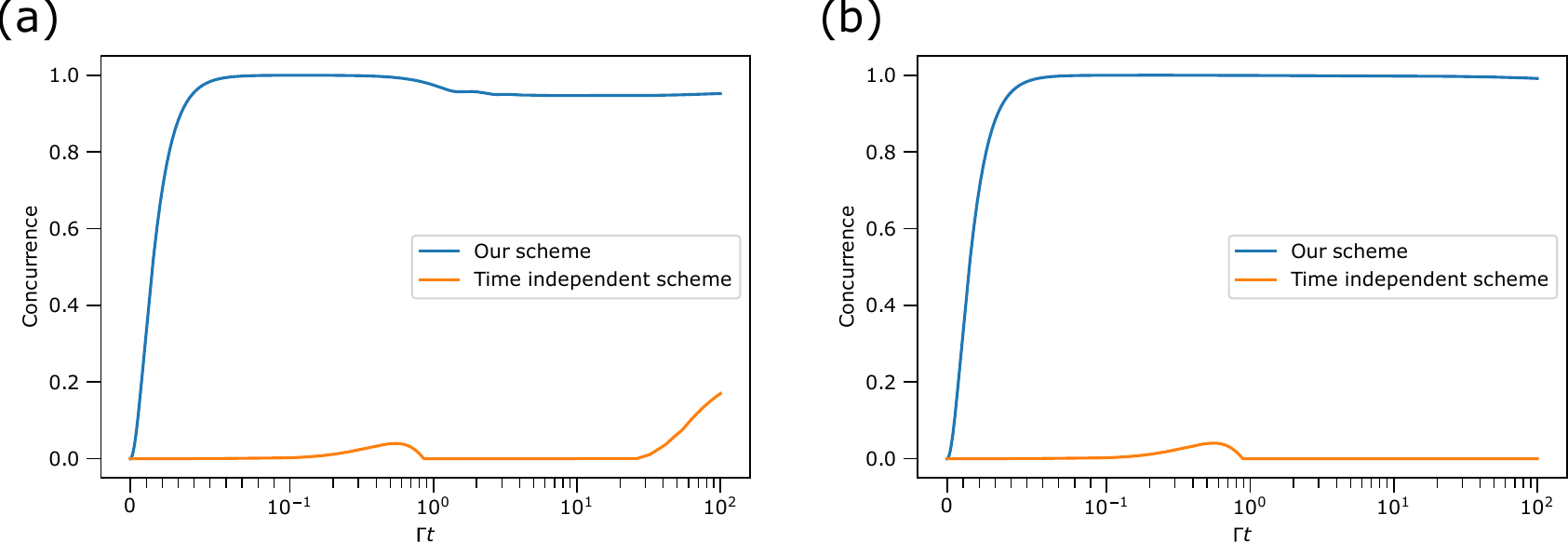}
  \caption[]{
  In (a) and (b) respectively, we plot the effect of detunings on
  the average concurrence when $N=8$ qubits form $N/2=4$ dimer pairs for the cases when the
  detunings on each qubit obey \eqref{appendix eqn: Zoller detuning
  pattern dimers} and when the detunings on each qubit do not obey \eqref{appendix eqn: Zoller detuning pattern dimers}.
  In (a), we see our scheme quickly reaches a high value of
  concurrence that decays to its true steady state value, which is lesser
  than $1$ because of the detuning. On the other hand, the time independent
  scheme sees a rise in concurrence only after a long time.
  This shows that when \eqref{appendix eqn: Zoller detuning pattern dimers} is
  satisfied for the detuning patterns, both our scheme and the time
  independent scheme can generate dimerised pairs in the steady state, though
  our scheme accelerates the approach to this steady state.
  In (b), we see that while a violation of \eqref{appendix eqn: Zoller detuning pattern dimers} means that we will not get dimerised pairs in the steady state, our scheme is able to rapidly produce long-lived concurrence that decays quite slowly for the values of $\delta_i$ considered. On the other hand, the time-independent scheme fails to produce any meaningful amount of concurrence. The numerical parameters chosen are: $\Omega_{\text{max}}/\Gamma = 5$, $\delta_i/\Gamma = \{0.25, -0.25, 0.275, -0.275, 0.3, -0.3, 0.325, -0.325\}$ for (a) and $\delta_i/\Gamma = \{0.300, 0.296, 0.426, 0.405, 0.381, 0.459, 0.292, 0.429\}$ for (b), $\Delta \gamma = 0$. We also use \eqref{appendix eqn: function for theta} in our scheme, with $k = 10$. Lastly, for our scheme, we set $\Omega(t)$ to increase linearly from $0$ to its max value $\Omega_{\text{max}}$ after $\Gamma t =1$, after which it will stay constant.
  }
  \label{fig: effect of detunings}
\end{figure}
In the event that $\delta_i \neq 0$ and that $\delta_i$ does not obey the
detuning pattern in \eqref{appendix eqn: Zoller detuning pattern dimers}, then the
state of dimerised pairs will not be the steady state of the system. Thus,
while our protocol can still produce the state of dimerised pairs, the $i$th
dimerised pair decays with a rate proportional to $\delta_i/\Omega$. On the
other hand, since $\delta_i$ does not follow the detuning pattern in
\eqref{appendix eqn: Zoller detuning pattern dimers}, the time independent scheme will
not produce the state of dimerised pairs regardless of how long one waits,
since the dimerised pair state will not be the steady state of the system.
Since it is generally experimentally easier to set up detuning patterns
$\delta_i$ that will not obey \eqref{appendix eqn: Zoller detuning pattern dimers},
the fact that our scheme works in this case but the time independent scheme
does not is noteworthy. A numerical example comparing our scheme to the
time-independent scheme in this case is shown in Figure~\ref{fig: effect of detunings}.}

\clearpage

\section{$N=6$ multimers}
\label{appendix: N=6 multimer}
First, we show numerically that our scheme works for a $N=6$ multimer, i.e
when our target state is
\begin{equation}
  \ket{\Phi} = \sum \ket{S}_{i_1,i_2}\ket{S}_{i_3,i_4}\ket{S}_{i_5,i_6}
\end{equation}
where the summation is over all possible pairs
$\{(i_1,i_2),(i_3,i_4),(i_5,i_6)\}$ where $i_k < i_{k+1}$. By counting, we
see that for $N$ spins, we would have $(N-1)(N-3)\dots 1$ terms in the
summation. Hence, for $N = 6$, this gives us $15$ terms in our summation.
Using $\Omega(t) = mt, m>0$ and \eqref{appendix eqn: function for theta} for
$\theta(\Omega(t))$, in the case where $\Delta \gamma = 0$, a straightforward
application of our scheme gives us the results in \figref{fig: N=6 multimer
result}. Clearly, we are able to easily obtain the state $\ket{\Phi}$
numerically, and we are able to do so in $\Gamma t \ll 1$. However, for this
case, the $X$ operator is a linear combination of multiple many-body
interaction terms, which means that experimental implementation of this
scheme is still quite tricky with the current state of quantum control.

\begin{figure}[H]
  \centering
  \includegraphics[width=0.45\textwidth, trim={0cm 0cm 0cm 0cm},clip]{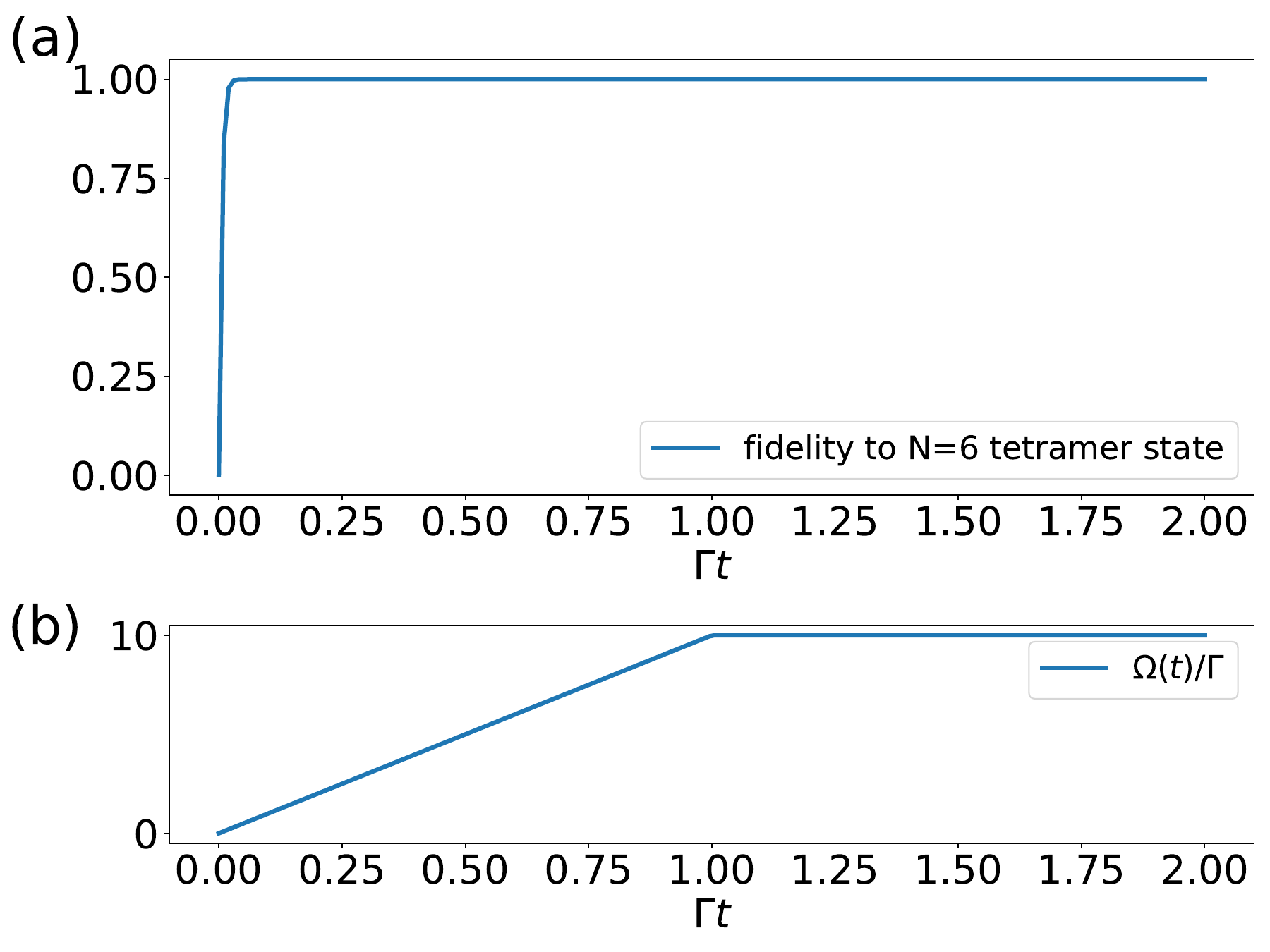}
  \caption[]{In (a), we plot the fidelity of the system state to
  $\ket{\Phi}$, and in (b), we plot the time variation of the driving
  strength $\Omega(t)$. The numerical results here show that at least in
  theory, our scheme works.}
  \label{fig: N=6 multimer result}
\end{figure}

\end{document}